\documentclass[journal]{IEEEtran}
\usepackage{balance}
\usepackage[a4paper, left=1.27cm, right=1.27cm, top=1.815cm, bottom=4.3cm]{geometry}

\usepackage{float}
\usepackage{standard}
\usepackage{etex}

\usepackage{rotate}
\usepackage{algorithmic}
\usepackage{algorithm}
\usepackage{float}
\usepackage{tikz}
\usepackage{refcount}

\usepackage{pgf,tikz}
\usepackage{pgfplots}
\usepackage{pgfplotstable}
\usepackage{bm}
\usepackage{adjustbox}
\usetikzlibrary{calc}
\usepackage{relsize}
\usepackage{ifthen}

\usetikzlibrary{calc,fit,arrows,plotmarks,intersections,patterns,shapes,decorations,positioning, backgrounds, matrix}

\usepackage[numbers,sort&compress]{natbib}
\usepackage[ngerman, english]{babel}

\usetikzlibrary{arrows.meta}
\edef\wdArrowLength{2}
\tikzset{>={Latex[width=1.5mm,length=\wdArrowLength mm]}}
\usepackage{graphicx}
\usepackage[outdir=./pictures]{epstopdf}

\usepackage{cancel}
\usepackage{mathtools}
\usepackage{shadethm}
\usepackage{empheq}
\usepackage{skull}
\usepgfplotslibrary{units}
\usepackage{calc}
\usepackage{nicefrac}
\usepackage{fancybox}
\usepackage{psfrag}
\usepackage{dsfont}

\title{From Metric to Mechanism: Designing Wireless Resilience through Finite Blocklength Dynamics}
\author{\IEEEauthorblockN{Kevin Weinberger\IEEEauthorrefmark{1}, Aydin Sezgin\IEEEauthorrefmark{1} and  Mehdi Bennis\IEEEauthorrefmark{2}\\}
\IEEEauthorblockA{\IEEEauthorrefmark{1}Institute of Digital Communication Systems, Ruhr University Bochum, Germany\\
\IEEEauthorrefmark{2}Centre for Wireless Communications, University of Oulu, Finland \\
Email: {\{kevin.weinberger,aydin.sezgin\}@rub.de}, mehdi.bennis@oulu.fi }
\thanks{This work was supported in part by the German Research Foundation (DFG) in the course of the project SPP2433 under the project no. 541021107 (Measurement Technology on Flying Platforms) under grant SE 1697/22-1.}
}
\date{\today}

\usepackage{graphicx}
\usepackage{placeins}

\usepackage{booktabs}
\usepackage{marvosym}
\usepackage{multirow}

\usepackage{latexsym, amsmath, amssymb, amsfonts, upgreek}

\usepackage[nolist]{acronym}

\usepackage{tikz}
\usetikzlibrary{calc,arrows,positioning,decorations,shadows,shapes,fadings,matrix}
\tikzset{>=latex'}
\tikzset{semithick}

\makeatletter
\providecommand{\IfElsePackageLoaded}[3]{\@ifpackageloaded{#1}{#2}{#3}}
\makeatother
\usepackage{graphicx}
\usepackage{subcaption}

\makeatletter

\def\tikz@delimiter#1#2#3#4#5#6#7#8{%
	\bgroup
		\pgfextra{\let\tikz@save@last@fig@name=\tikz@last@fig@name}%
		node[outer sep=0pt,inner sep=0pt,draw=none,fill=none,anchor=#1,at=(\tikz@last@fig@name.#2),#3]
		{%
			{\nullfont\pgf@process{\pgfpointdiff{\pgfpointanchor{\tikz@last@fig@name}{#4}}{\pgfpointanchor{\tikz@last@fig@name}{#5}}}}%
			\delimitershortfall\z@
			\resizebox*{!}{#8}{$\left#6\vcenter{\hrule height .5#8 depth .5#8 width0pt}\right#7$}%
		}
		\pgfextra{\global\let\tikz@last@fig@name=\tikz@save@last@fig@name}%
	\egroup%
}

\tikzset{hexagon/.code={
	\draw (0,2) -- (-4,0) -- (0,-2) -- (4,0) -- (0,2);
}}

\tikzset{phone/.code={
   \node [rectangle,rounded corners=1.5pt,draw,minimum height=0.6cm, minimum width=0.35cm] at (0,0){};
   \node [rectangle,rounded corners=1.5pt,draw,minimum height=0.5cm, minimum width=0.3cm] at (0,0){};
}}

\makeatletter
\def\cantox@vector#1#2#3#4#5#6#7#8{%
  \dimen@.5\p@
  \setbox\z@\vbox{\boxmaxdepth.5\p@
   \hbox{\kern-1.2\p@\kern#1\dimen@$#7{#8}\m@th$}}%
  \ifx\canto@fil\hidewidth  \wd\z@\z@ \else \kern-#6\unitlength \fi
  \ooalign{%
    \canto@fil$\m@th \CancelColor
    \vcenter{\hbox{\dimen@#6\unitlength \kern\dimen@
      \multiply\dimen@#4\divide\dimen@#3 \vrule\@depth\dimen@\@width\z@
      \vector(#3,-#4){#5}%
    }}_{\raise-#2\dimen@\copy\z@\kern-\scriptspace}$%
    \canto@fil \cr
    \hfil \box\@tempboxa \kern\wd\z@ \hfil \cr}}
\def\bcancelto#1#2{\let\canto@vector\cantox@vector\cancelto{#1}{#2}}
\makeatother

\IEEEoverridecommandlockouts
\usepackage{algorithmic}
\begin{document}
\maketitle
\begin{abstract}
Next-generation wireless networks must maintain reliable operation under abrupt and severe disruptions, particularly in ultra-reliable low-latency communication (URLLC) scenarios where strict time constraints dominate system design. This work addresses network resilience from a time-centric perspective by explicitly integrating finite blocklength (FBL) communication, thereby exposing transmission duration as a controllable resource for system recovery. To this end, we propose a unified cross-layer framework that jointly couples queue dynamics, rate adaptation, and blocklength optimization, enabling the system to actively absorb, adapt to, and recover from diverse resilience events.

To systematically evaluate these mechanisms, we introduce an interpretable resilience metric that decomposes disruption impact into absorption loss, adaptation efficiency, and recovery behavior, enabling a direct and intuitive assessment of system resilience. Building on this framework, we develop a three-stage alternating optimization approach that jointly optimizes PHY-layer parameters, including beamforming, reconfigurable intelligent surface (RIS) phase shifts, and blocklength, revealing the importance of time-aware resource allocation in the FBL regime. Numerical results demonstrate strong resilience performance under repeated channel disruptions and AI-driven traffic surges, highlighting the effectiveness of cross-layer resource adaptation. Finally, the proposed resilience metric enables an intuitive and consistent comparison of resilience performance across different approaches and disruption types, while revealing their respective strengths and limitations.
\end{abstract}

\thispagestyle{empty}
\pagestyle{empty}

\section{Introduction}
\subsection{Motivation}
As the vision of \ac{6G} networks emerges, future wireless systems are expected to support not only human-centric applications but also mission-critical services with stringent \ac{URLLC} requirements~\cite{you2021towards}. Applications such as autonomous driving, industrial automation, and remote medical systems impose strict \ac{QoS} constraints, where even short disruptions can lead to significantly impacted safety and reliability~\cite{ChaccourAOI}. In many cases, satisfying these requirements is becoming feasible only through \ac{AI}-driven techniques, which in turn further reshape network load characteristics and communication requirements \cite{SemCom}.
To address these challenges, next-generation systems must move beyond traditional notions of robustness and instead embrace resilience, defined as the ability to absorb unexpected disturbances, adapt to changes in the network dynamically, and recover rapidly with minimal performance loss~\cite{brosInArms,RobertRes,ResByDesign}. Especially within \ac{URLLC} scenarios, these requirements are further intensified by tight temporal constraints, where recovery must occur within very short transmission intervals \cite{URLLC_Res}.

These stringent delay and reliability constraints fundamentally limit the blocklength available for communication and thus places the system in the \ac{FBL} regime, where reliability–rate–latency trade-offs become fundamental. Unlike the classical \ac{IBL} setting, \ac{FBL} coding introduces both a rate penalty and a non-zero decoding error probability due to the use of short codewords. Consequently, recovery becomes less efficient, as successfully delivered throughput is reduced not only by stronger channel coding but also by retransmissions or packet losses caused by decoding errors \cite{weinberger2025symbolcountsresilientwireless}. If this service deficit persists, packets accumulate in the transmit queues and may eventually be dropped due to buffer overflow. Therefore, transmit queues at the link layer play a central role by decoupling traffic generation from transmission and smoothing bursty arrivals under time-varying channel conditions.


This buffering capability becomes particularly relevant in \ac{AI}-driven wireless systems, where traffic generation is tightly coupled to task execution and learning-based demand patterns. These workloads are highly dynamic, event-driven, and prone to abrupt surges \cite{GenAI}, creating rapid variations in the required service rate and increasing the likelihood of temporary traffic-service mismatches. Consequently, transmit queues no longer serve solely as congestion control mechanisms but become essential resilience enablers that can absorb short-term demand fluctuations and maintain service continuity under disruptive conditions \cite{STERBENZ20101245}.


Building on these insights, we characterize resilience through a two-phase structure consisting of an absorption phase and an adaptation phase ~\cite{resiliencemetric,RobertRes}. The latter explicitly accounts for performance deficits incurred during the disturbance and compensates for them within the same coherence block, thereby mitigating long-term performance degradation. To systematically capture this behavior, we introduce a unified resilience metric that jointly evaluates absorption, adaptation and \ac{TTR} performance while explicitly accounting for performance loss during disruptions. Virtual queues are employed to encode deviations from desired performance, thereby embedding backlog evolution into the system state and enabling a holistic assessment of resilience dynamics.

This formulation enables compensation within a finite coherence block, where accumulated deficits are actively recovered to restore both queue stability and nominal operation. From a queueing perspective, resilience is therefore not limited to restoring nominal system performance, but also requires draining accumulated backlog to recover the pre-disturbance state \cite{Joint_BL_BF}. This temporal recovery process naturally exposes additional design degrees of freedom. In particular, treating the blocklength as a controllable parameter allows the system to shape the temporal structure of recovery, since the timeslot duration directly depends on the blocklength under fixed bandwidth \cite{FBL_chanuse}.

This flexibility elevates resilience from a purely reactive mechanism to a system-level decision problem, in which the network must balance the urgency of immediate recovery against the cost of operating under \ac{FBL} constraints, determining whether rapid compensation or delayed adaptation yields the best overall performance. This interplay between strict temporal constraints and performance degradation under \ac{FBL} regimes necessitates physical-layer mechanisms that can react to environmental variations in real time.

Rather than relying on a single adaptation strategy, resilience can be enhanced by jointly exploiting multiple physical-layer degrees of freedom, each addressing a different aspect of the communication process.
Among these, transmit beamforming enables the network to spatially redistribute available resources toward affected users, improving link quality through adaptive signal steering. In parallel, \acp{RIS} provide a software-controlled propagation environment by dynamically reconfiguring electromagnetic reflections, creating additional spatial diversity and improving channel conditions without increasing transmit power \cite{basar2019wireless, SynBenefits}. Finally, the blocklength itself represents a critical adaptation variable in the \ac{FBL} regime, since it directly governs the trade-off between latency, reliability, and achievable rate. While shorter blocklengths enable faster reactions to disturbances, they also incur a finite-blocklength penalty that can substantially degrade throughput \cite{weinberger2025symbolcountsresilientwireless}.
Consequently, resilience under stringent latency requirements cannot be understood by considering these mechanisms in isolation. Instead, it depends on the coordinated adaptation of beamforming, \ac{RIS} configuration, and blocklength allocation, whose interactions determine both the immediate response to disruptions and the subsequent recovery process. Capturing these interactions therefore requires a resilience framework that explicitly accounts for both the temporal characteristics of \ac{FBL} communications and the available physical-layer adaptation mechanisms.
To investigate this challenge, we develop a resilience framework that explicitly incorporates \ac{FBL} effects during the system's recovery phase. The utilized model quantifies resilience in terms of three core capabilities: absorption, adaptation, and Relative Recovery Index (RRI), allowing us to study the performance trade-offs between reacting to an outage or continuing operation with limited resources. Further, we study the proposed resilience framework in terms of system throughput, enabling a quantitative assessment of how absorption, adaptation, and recovery dynamics jointly affect end-to-end performance under \ac{FBL} constraints.

Based on this framework, we develop an adaptive optimization algorithm that dynamically allocates blocklengths during the recovery process and evaluate its performance using the proposed resilience metric. The results demonstrate that dynamic blocklength adaptation improves recovery efficiency across multiple disruption events, enabling faster and more effective restoration of system performance. Furthermore, the proposed metric provides an intuitive and consistent evaluation of resilience behavior, allowing direct comparison of different disruption scenarios and recovery strategies.

\subsection{Related Literature}
Wireless-network resilience has been studied through a mix of service-oriented frameworks, survivability indices, and attack-specific metrics rather than a single universal definition. In \cite{RobertRes}, resilience is addressed in the context of mixed-criticality systems, where a joint metric captures performance degradation and recovery under outage conditions within a rate-splitting-based resource allocation framework. A broader perspective is provided in \cite{ResByDesign}, which introduces a resilient-by-design paradigm based on prediction, preemption, protection, and progression, emphasizing resilience as a system-level system property rather than a single metric. More recently, \cite{Dressler_Res} highlights the role of cross-technology communication in improving resilience through increased path diversity, while \cite{bennis2025resilient} argues for a clearer mathematical separation between resilience, robustness, and reliability.

Complementing these system-level views, recent advances in URLLC have emphasized the importance of finite blocklength (FBL) constraints and cross-layer optimization for achieving stringent latency and reliability requirements. Foundational work in \cite{URLLC-FBL} developed cross-layer formulations that jointly consider queueing and transmission delays under FBL constraints, including proactive packet dropping to maintain QoS. Building on this, \cite{Joint_BL_BF} proposed path-following algorithms for joint blocklength and beamforming optimization in short-packet regimes, while \cite{URLLC_RISOPT} and \cite{RIS_RES} demonstrated that RIS-assisted communication can significantly improve FBL performance and mitigate outage-induced degradation. Further extensions in \cite{Joint_RIS_FBL_BF} highlight additional gains from jointly optimizing blocklength, beamforming, and RIS configuration, and \cite{weinberger2025symbolcountsresilientwireless} identifies critical blocklength thresholds for reliable recovery under FBL constraints.

However, existing work often studies resilience from separate angles, such as queue dynamics, beamforming design, blocklength adaptation, or RIS optimization, rather than considering them jointly in a unified framework. Because of this separation, transmission time is rarely treated as a controllable resource for system recovery, and existing methods do not provide interpretable metrics that describe resilience across absorption, adaptation, and recovery phases.
In contrast, this work introduces a unified and interpretable cross-layer resilience metric for wireless systems, explicitly capturing the interaction between traffic dynamics, service rates, and recovery behavior.
\subsection{Contributions}
The main contributions of this work are summarized as follows:
%
%
%
%
\begin{itemize}
    \item \textbf{Resilience metric:} We propose an interpretable virtual-queue-based resilience metric that captures temporal performance deviations through cumulative absorption, adaptation, and recovery effects, enabling a unified and extensible assessment of resilience behavior.

    \item \textbf{\ac{FBL}-aware resilience analysis:} We incorporate finite blocklength effects into resilience modeling and identify blocklength adaptation as an additional temporal resource for controlling the trade-off between immediate absorption and long-term recovery.

    \item \textbf{Two-phase resilience framework and optimization:} We develop a joint absorption--adaptation framework that explicitly models disruption response and recovery under finite blocklength constraints. The framework enables adaptive resource allocation through backlog-aware scheduling, beamforming, RIS phase-shift optimization, and dynamic blocklength adaptation to maintain queue stability under severe disruptions.

    \item \textbf{Performance evaluation:} We demonstrate the effectiveness of the proposed framework through numerical evaluations under repeated channel outages and AI-driven traffic surges, showing improved resilience performance and enabling intuitive comparison of different disruption scenarios through the proposed metric.
\end{itemize}

\section{System Model}\label{ch:Sysmod}
\begin{figure*}
	\centering \includegraphics[width=1\linewidth , trim=5 20 25 5,
    clip]{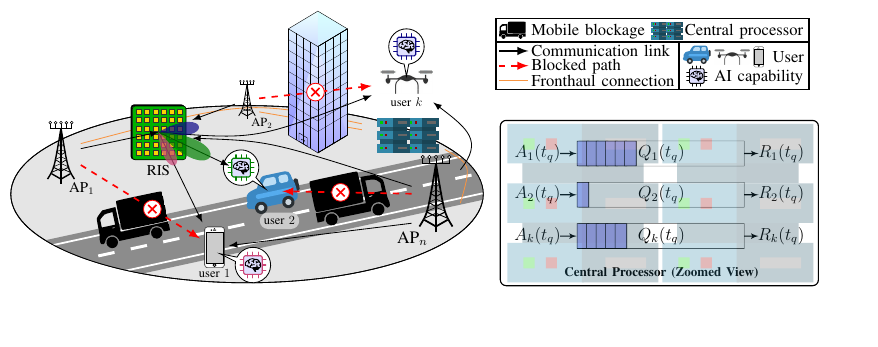}
	\caption{System model of a multi-AP communication network with AI-driven traffic and dynamic blockage mitigation. Mobile and static blockages may disrupt direct AP-user links, motivating adaptive traffic rerouting via RISs or alternative APs. The APs are coordinated by a CP with per-user transmit queues, where the queue dynamics are described by the queue state $Q_k(t_q)$, arrival rate $A_k(t_q)$, and departure rate $R_k(t_q)$ at time instance $t_q$. The inset provides a detailed view of the CP queuing model.}
	\label{fig:1}
\end{figure*}

This work investigates a cell-free \ac{MIMO} downlink system enhanced by a \ac{RIS}, as illustrated in Fig.~\ref{fig:1}. Specifically, a group of single-antenna users $\mathcal{K} = \{1, \cdots, K\}$ is served by multiple \acp{AP}, each equipped with $L$ antennas, indexed by the set $\mathcal{N} = \{1, \cdots, N\}$. To introduce additional propagation paths, an $M$-element \ac{RIS}, arranged in a uniform planar array, is deployed within the center of the served area. Its placement is carefully selected to ensure that it can establish alternative links in the event that the direct AP-user channels become obstructed. The \ac{RIS} and all \acp{AP} are connected to a \ac{CP} via orthogonal fronthaul connections. As illustrated in Fig.~\ref{fig:1}, with a zoomed-in view of shown on the right side, data packets for each user arrive at the \ac{CP} and are assigned to the corresponding user queue $Q_k(t_q)$ at time instant $t_q$. Each queue is characterized by the number of arriving packets $A_k(t_q)$ within each time slot. To satisfy the user requirements, the queues must remain stable, i.e., the long-term service rate must match the average traffic demand. The instantaneous service rate provided to user $k$ at time instant $t_q$ is denoted by $R_k(t_q)$ and determines the evolution of the corresponding queue state.


\subsection{Channel Model}\label{ssec:chan}
\newcommand{\RisChanFull}[1]{\vect{h}_{#1} + \mat{G}_{#1}\vect{v}}
\newcommand{\RisChan}[1]{\vect{h}^\mathsf{eff}_{#1}}
\subsubsection{Perfect CSI}
The channel model adopts a quasi-static memory-less block fading assumption, where the channel coefficients remain constant over a coherence interval of duration $T_\mathsf{coh}$ and vary independently between different coherence blocks. The transmit signal vector from the $n$-th \ac{AP} is defined as
\[
\vect{x}_{n} = \sum_{k \in \mathcal{K}} \vect{w}_{n,k} s_k, \quad \forall n \in \mathcal{N},
\]
where $s_k$ denotes the symbol intended for user $k$, drawn from an i.i.d. Gaussian codebook, and $\vect{w}_{n,k} \in \mathbb{C}^{L \times 1}$ is the beamforming vector provided by the central processor (\ac{CP}). The per-AP transmit signal is constrained by a maximum power limit:
\begin{align}\label{eq:powConst}
\mathbb{E}\{\vect{x}_n^H \vect{x}_n\} = \sum_{k \in \mathcal{K}} \| \vect{w}_{n,k} \|_2^2 \leq P^{\mathsf{max}}_n, \quad \forall n \in \mathcal{N}.
\end{align}

The aggregate transmit signal vector across all \acp{AP} can then be expressed as $
\vect{x} = [\vect{x}_1^T, \vect{x}_2^T, \dots, \vect{x}_N^T]^T \in \mathbb{C}^{NL \times 1}.
$
The direct channel between AP $n$ and user $k$ is denoted as $\vect{h}_{n,k} \in \mathbb{C}^{L \times 1}$. For RIS-assisted communication, the reflected channel is modeled as
\[
\mat{G}_{n,k} = \mat{H}_{n} \mathsf{diag}(\vect{v}) \vect{g}_k \in \mathbb{C}^{L \times 1},
\]
where $\mat{H}_n \in \mathbb{C}^{L \times M}$ is the AP-to-RIS channel, $\vect{g}_k \in \mathbb{C}^{M \times 1}$ is the RIS-to-user channel, and $\mathsf{diag}(\vect{v})$ contains the RIS reflection coefficients $\vect{v} = [v_1, v_2, \dots, v_M]^T$, with each $v_m = e^{j\theta_m}$ and $\theta_m \in [0, 2\pi]$. We define the aggregated direct channel vector for user $k$ as
$
\vect{h}_k = [\vect{h}_{1,k}^T, \vect{h}_{2,k}^T, \dots, \vect{h}_{N,k}^T]^T \in \mathbb{C}^{NL \times 1},
$
and the full AP-to-RIS channel matrix as
$
\vect{H} = [\mat{H}_1^T, \mat{H}_2^T, \dots, \mat{H}_N^T]^T \in \mathbb{C}^{NL \times M}.
$

Using these definitions, the effective received signal at user $k$ is given by the superposition of direct and RIS-reflected paths:
\begin{align}\label{recSgn}
y_k = (\vect{h}_k + \mat{G}_k \vect{v})^H \vect{x} + n_k = (\vect{h}_k^\mathsf{eff})^H \vect{x} + n_k,
\end{align}
where $\mat{G}_k = \vect{H} \mathsf{diag}(\vect{g}_k)$, $\vect{h}_k^\mathsf{eff} = \vect{h}_k + \mat{G}_k \vect{v}$, and $n_k \sim \mathcal{CN}(0, \sigma_k)$ is complex \ac{AWGN}.

\noindent
The received signal (\ref{recSgn}) at user $k$ is then given by
\begin{align}
y_k = (\RisChan{k})^H \vect{w}_k s_k + \sum_{i\in\mathcal{K}\setminus\{k\}} (\RisChan{k}) \vect{w}_i s_i + n_k,
\end{align}
where the first term represents user $k$'s desired signal, and the second term accounts for all the other user's interference.

\subsubsection{Imperfect CSI}
In practice, the \ac{CP} has access only to imperfect \ac{CSI}. To enable beamforming design, the direct and RIS-assisted channel components are estimated through uplink pilot signaling. Following the \ac{MMSE}-based channel estimation framework of \cite{channelEst,SynBenefits}, the channels are modeled as
\begin{align}
\hat{\vect h}_k &= \vect h_k + \Delta\vect h_k,\\
\hat{\mat G}_k &= \mat G_k + \Delta\mat G_k,
\end{align}
where $\hat{\vect h}_k$ and $\hat{\mat G}_k$ denote the \ac{MMSE} estimates, while $\Delta\vect h_k$ and $\Delta\mat G_k$ represent the corresponding estimation errors. The estimated effective channel can therefore be expressed as
\begin{align}
\hat{\vect h}_k^\mathsf{eff}
&=
({\vect h}_k+\Delta\vect h_k)
+
({\mat G}_k+\Delta\mat G_k)\vect v
\end{align}
By the orthogonality principle of \ac{MMSE} estimation, the channel estimates and estimation errors are mutually uncorrelated. The corresponding error covariance matrices depend on the pilot duration, pilot transmit power, receiver noise variance, and large-scale fading coefficients, and are given in \cite{SynBenefits}.
\subsubsection{Infinite Blocklength Regime}
With the above expressions, we formulate the \ac{SINR} of user $k$ decoding its message as
\begin{align}
\Gamma_k = \frac{|(\RisChan{k})^H \vect{w}_k|^2}{\sum_{i\in\mathcal{K}\setminus\{k\}}|(\RisChan{k})^H \vect{w}_i|^2 + \sigma^2},
\end{align}
where $\sigma^2$ denotes the noise power. Consequently, each user can successfully decode its message, if the following condition holds:
\begin{align}
C_k \leq B \log_2(1+\Gamma_k) ,
\end{align}
where $B$ denotes the transmission bandwidth and $C_k$ represents the allocated rate of user $k$ under \ac{IBL} assumption.
\subsubsection{Finite Blocklength Regime}
In case of a disruption, however, the system needs to reallocate resources quickly and consequently operates with \acp{FBL} of size $\eta$. According to \cite{FBL_Polyanskiy}, user $k$'s QoS demands in the \ac{FBL} regime are satisfied, if the following condition holds
\begin{align}\label{FBL_rate}
R_k(\vect{w},\vect{v},\eta) \leq C_k -  B\Omega \, \sqrt{{V_\mathsf{disp}(\Gamma_k)}/{\eta}} ,
\end{align}
where $R_k$ is the allocated rate for user $k$, $\Omega = \mathcal{Q}^{-1}(\epsilon) \log_2(e)$, $\eta$ is the blocklength, $\epsilon$ is the \ac{BLER} and the function $\mathcal{Q}^{-1}(\cdot)$ represents the inverse of the Gaussian $\mathcal{Q}$ function\footnote{$\mathcal{Q}(x) = \int_{x}^{\infty} \frac{1}{\sqrt{2\pi}}\exp{(\frac{t^2}{2})}dt.$}. Moreover, $V(\Gamma_k)$ is the channel dispersion parameter given by \begin{align}\label{FBL_disp}
  V_\mathsf{disp}(\Gamma_k) = 1 - (1+\Gamma_k)^{-2},
\end{align}
which is justified under the assumption of Gaussian signaling in our model. The penalty term in \eqref{FBL_rate} captures the finite blocklength effect, which reduces the achievable rate to maintain the target \ac{BLER} $\epsilon$ under short packet transmission. Because this term depends on the blocklength $\eta$, it couples reliability and latency. As a result, the transmission duration varies with $\eta$, inducing a variable transmission time structure, which is formalized in the following subsection.
\newcommand{\RisOnlyChan}[1]{\vect{h}_{#1}^\mathsf{RIS}}

%
%
%
%

\subsection{Queuing Model}

We adopt a queuing model at the \ac{CP} that acts as the transmit buffer for the data destined for the users in the network. Each user $k\in\mathcal{K}$ is assigned a queue $Q_k^\mathsf{buff}$. Packet arrivals are modeled as a Poisson process with nominal arrival intensity $\xi_k$ (packets/s). Consequently, the number of packet arrivals during time slot $q$, denoted by $A_k(t_q)$, follows
\begin{align}
A_k(t_q)\sim\mathrm{Poisson}(\xi_kT_q),
\end{align}
with mean
\begin{align}
\bar{A}_k(t_q)\triangleq\mathbb{E}[A_k(t_q)]=\xi_kT_q.
\end{align}

At time instance $t_q$, the queue dynamics are represented in a bit-wise fashion as
\begin{align}\label{eq:Q_base}
Q_k^\mathsf{buff}(t_{q+1})
=
\left[
Q_k^\mathsf{buff}(t_q)
-
T_qR_k(t_q)
+
M_k^p(t_q)A_k(t_q)
\right]^+,
\end{align}
where $A_k(t_q)$ denotes the random number of packet arrivals during slot $q$, $[x]^+=\max\{0,x\}$, and $M_k^p(t_q)$ denotes the random packet size (in bits) of user $k$. To capture bursty traffic and heterogeneous packet formats, the packet size is allowed to vary over time with nominal mean
\begin{align}
\bar{M}_k^p\triangleq\mathbb{E}[M_k^p(t_q)].
\end{align}
Hence, the total number of arriving bits during slot $q$ is given by the product $M_k^p(t_q)A_k(t_q)$. Moreover, the departure rate $R_k(t_q)$ denotes the service rate in bits/s, such that $T_qR_k(t_q)$ represents the number of successfully transmitted bits during slot $q$, while $\Delta Q_k^\mathsf{buff}(t_q)$ denotes the corresponding queue variation.

To prevent buffer overflow and the loss of potentially critical information, the transmission scheme must ensure that all queues remain stable. A queue is considered stable if
\begin{align}\label{Q:steady}
\limsup_{t\rightarrow\infty}
\frac{1}{t}
\sum_{\tau=0}^{t-1}
\mathbb{E}[Q_k^\mathsf{buff}(\tau)]
<
\infty.
\end{align}
Consequently, queue stability is achieved when the long-term average service rate exceeds the long-term average arrival bit rate \cite{neely2010introduction}, i.e.,
\begin{align}\label{MeanRate}
\bar{R}_k
\ge
\mathbb{E}\!\left[M_k^p(t_q)A_k(t_q)\right],
\qquad
\forall k\in\mathcal{K}.
\end{align}
Under the common assumption that packet arrivals and packet sizes are statistically independent, the stability condition simplifies to
\begin{align}
\bar{R}_k
\ge
\bar{M}_k^p\,\bar{A}_k(t_q)
=
\bar{M}_k^p\,\xi_kT_q,
\qquad
\forall k\in\mathcal{K}.
\end{align}

\section{Virtual-Queue-Based Resilience under Finite Blocklength Constraints}

This section introduces a variable time-slot structure in the \ac{FBL} regime as a key mechanism for enabling time-aware resilience. The main focus is on effective throughput in \ac{URLLC} systems, although the presented framework is generalizable to other performance metrics such as energy efficiency, delay, or reliability. Building on this, we develop a virtual-queue-based representation of the system, which allows queue dynamics and service constraints to be expressed in a tractable form that directly captures resilience behavior. In particular, the virtual queue represents the accumulated performance loss that must be compensated in future transmissions to maintain the system at its nominal operating point. Within this framework, we distinguish between distinct resilience phases, namely absorption and adaptation, which describe how the system responds to and recovers from disruptions over time. Finally, we introduce an interpretable resilience metric that quantifies system performance across these phases and explicitly reflects the interplay between blocklength selection, transmission rates, and virtual queue evolution.

\subsection{Variable Time Slot Structure in the FBL Regime}
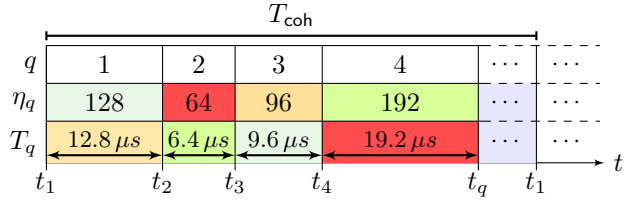
\begin{figure}
  \centering

\begin{tikzpicture}[scale=1,xscale=1.2]

\draw[->] (0,0) -- (6.15,0) node[right] {$t$};
\draw[-] (0,0) -- (0,1.55);


\draw (0,0.25) node[left] {$T_q$};
\draw (0,0.8) node[left] {$\eta_q$};
\draw (0,1.3) node[left] {$q$};

\edef\xValI{1.28}
\edef\xValII{\xValI+0.8}
\edef\xValIII{\xValII+0.96}
\edef\xValIV{\xValIII+1.72}
\edef\xValV{\xValIV+0.64}
\edef\xValVI{\xValV+0.7}

\fill[green!70!black!10]
(0,1.05) rectangle (\xValI,0.55);

\fill[red!70]
(\xValI,1.05) rectangle (\xValII,0.55);

\fill[orange!60!yellow!40]
(\xValII,1.05) rectangle (\xValIII,0.55);

\fill[green!40!yellow!40]
(\xValIII,1.05) rectangle (\xValIV,0.55);

\fill[blue!10] (\xValIV,1.05) rectangle (\xValV,0.55);

\fill[yellow!60!orange!40]
(0,0) rectangle (\xValI,0.55);

\fill[green!40!yellow!40]
(\xValI,0) rectangle (\xValII,0.55);

\fill[green!70!black!10]
(\xValII,0) rectangle (\xValIII,0.55);

\fill[red!70]
(\xValIII,0) rectangle (\xValIV,0.55);

\fill[blue!10] (\xValIV,0) rectangle (\xValV,0.55);
\foreach \y in {0.55,1.05,1.55}{
    \draw (0,\y) -- ({\xValIV+0.05},\y);

    \draw[dashed] ({\xValIV+0.05},\y) -- ({\xValV-0.05},\y);

    \draw[dashed] (\xValV-0.05,\y) -- (6.1,\y);
}

\foreach \x/\name in {
0/t_1,
\xValI/t_2,
\xValII/t_3,
\xValIII/t_4,
\xValIV/t_q,
\xValV/t_1
}{
    \draw (\x,1.55) -- (\x,-0.1);
    \draw (\x,0) node[below] {$\name$};
}

\foreach \xStart/\xEnd in {
0/\xValI,
\xValI/\xValII,
\xValII/\xValIII,
\xValIII/\xValIV
}{
    \draw[<->,thick] (\xStart,0.1) -- (\xEnd,0.1);
}

\pgfmathsetmacro{\a}{(\xValI)/2}
\pgfmathsetmacro{\b}{(\xValI+\xValII)/2}
\pgfmathsetmacro{\c}{(\xValII+\xValIII)/2}
\pgfmathsetmacro{\d}{(\xValIII+\xValIV)/2}
\pgfmathsetmacro{\e}{(\xValIV+\xValV)/2}
\pgfmathsetmacro{\f}{(\xValV+\xValVI)/2}

\pgfmathsetmacro{\yA}{0.275} 
\pgfmathsetmacro{\yB}{0.8}   
\pgfmathsetmacro{\yC}{1.3}   

\node at (\a,\yA){\small $12.8\,\mu s$};
\node at (\b,\yA){\small$6.4\,\mu s$};
\node at (\c,\yA){\small$9.6\,\mu s$};
\node at (\d,\yA){\small$19.2\,\mu s$};
\node at (\e,\yA){\small$\cdots$};
\node at (\f,\yA){\small$\cdots$};
\node at (\a,\yB) {$128$};
\node at (\b,\yB) {$64$};
\node at (\c,\yB) {$96$};
\node at (\d,\yB) {$192$};
\node at (\e,\yB) {\small$\cdots$};
\node at (\f,\yB) {\small$\cdots$};
\node at (\a,\yC) {$1$};
\node at (\b,\yC) {$2$};
\node at (\c,\yC) {$3$};
\node at (\d,\yC) {$4$};
\node at (\e,\yC) {\small$\cdots$};
\node at (\f,\yC) {\small$\cdots$};
\draw[thick] (0,1.68) -- (\xValV,1.68)
    node[midway,above] {$T_\mathsf{coh}$};
\draw[thick] (0,1.62) -- (0,1.74);
\draw[thick] (\xValV,1.62) -- (\xValV,1.74);
\end{tikzpicture}
\caption{Variable-length time-slot scheduling model for $B = 10\,\mathrm{MHz}$ with non-uniform slot durations across $T_\mathsf{coh}$. Colors encode slot favorability w.r.t. latency ($T_q$) and rate penalty ($\eta_q$), i.e., shorter slots imply faster recovery but higher rate penalty.}\label{fig:flexTime}
\end{figure}

The utilized queuing model characterizes, how each users queue evolves over time based on arrival and departure processes. As the data is transmitted over the wireless network’s links, the departure rate depends on the channel conditions, the chosen transmission rate, the decoding success probability and the blocklength. Since the blocklength $\eta_q$ can vary from one time slot to another, the required duration $T_q$ for the transmission will vary as well. Motivated by this observation, we consider a variable-length time slot model, which allows us to represent the continuous-time evolution through a variable-length discrete-time process. More specifically, the duration $T_q$ required to transmit a codeword of length $\eta_q$ in time slot $q$ over a bandwidth-limited channel can be expressed as \cite{FBL_chanuse}:
\begin{align}\label{eq:TransTime}
T_q = \frac{\eta_q}{B}.
\end{align}
Thus, if transmission begins at time $t_q$, the next time slot starts at
\begin{align}
t_{q+1} = t_q + T_q = t_q + \eta_q/B, \quad q \in \mathbb{N},
\end{align}
which induces the non-uniform slot structure illustrated in Fig.~\ref{fig:flexTime}, where slot widths directly reflect the corresponding blocklengths $\eta_q$. We assume that each transmission is much shorter than the channel coherence time, i.e., $T_q \ll T_\mathsf{coh}$. Unlike fixed-length slot systems, this variable-length model captures the coupling between blocklength, transmission time, and bandwidth. In particular, the slot-dependent coloring in Fig.~\ref{fig:flexTime} highlights the relative favorability of each transmission with respect to latency ($T_q$) and rate penalty ($\eta_q$), where smaller slots result in faster system response at the cost of higher rate penalty. This temporal flexibility plays a key role in assessing the system’s resilience under varying blocklengths.

\subsection{Resilience with Virtual Queues}
In practical \ac{URLLC} systems, packet arrivals and departures may vary substantially across coherence periods. While moderate fluctuations are inherent to normal system operation, severe bursts or outage events are atypical and can cause rapid queue build-up, thereby requiring prompt system response. Consequently, using instantaneous queue variations directly to identify abnormal situations becomes unsuitable, as the scheduler would continuously react to normal traffic fluctuations and may therefore trigger overly aggressive control actions.

\begin{figure}
\begin{tikzpicture}[scale=1.2,xscale=1]

\draw[->] (0,0) -- (5.2,0) node[right] {$t$};
\draw[->] (0,0) -- (0,1.5);
\node at (-0.1,1.5)  {$Q_k^\text{(buff)}$};
\node[left] at (0,0) {\small 0};
\edef\xValI{1.28}
\edef\xValII{\xValI+0.95}
\edef\xValIII{\xValII+1.1}
\edef\xValIV{\xValIII+1.6}
\edef\xValV{\xValIV+0.64}
\edef\xValVI{\xValV+0.7}

\fill[green!70!black!10]
(0,0) rectangle (\xValI,1.3);

\fill[red!70]
(\xValI,0) rectangle (\xValII,1.3);

\fill[orange!60!yellow!40]
(\xValII,0) rectangle (\xValIII,1.3);

\fill[green!40!yellow!40]
(\xValIII,0) rectangle (\xValIV,1.3);

\node[scale=0.7, fill=white, fill opacity=0.7, text opacity=1, inner sep=2pt, rounded corners=2pt]
at ({0.5*\xValI},1.15)
{\small $R_k\!=\!\alpha^\mathsf{des}$ };

\node[scale=0.7, fill=white, fill opacity=0.7, text opacity=1, inner sep=2pt, rounded corners=2pt]
at ({0.5*(\xValI+\xValII)},1.15)
{\small $R_k \ll \alpha^\mathsf{des}$};

\node[scale=0.7, fill=white, fill opacity=0.7, text opacity=1, inner sep=2pt, rounded corners=2pt]
at ({0.5*(\xValII+\xValIII)},1.15)
{\small $R_k < \alpha^\mathsf{des}$};

\node[scale=0.7, fill=white, fill opacity=0.7, text opacity=1, inner sep=2pt, rounded corners=2pt]
at ({0.5*(\xValIII+\xValIV)},1.15)
{\small $R_k \gg \alpha^\mathsf{des}$};

\foreach \x/\name in {
0/t_1,
\xValI/t_2,
\xValII/t_3,
\xValIII/t_4,
\xValIV/t_q
}{
    \draw (\x,0) -- (\x,1.3);
    \node[below] at (\x,0) {$\name$};
}

\draw[thick,blue]
(0,0) -- (0,1);
\draw[thick,blue]
(0,1.0) -- (\xValI,0.0);
\draw[thick,blue]
(\xValI,0.0) -- (\xValI,0.40)
-- (\xValII,0.35);

\draw[thick,blue]
(\xValII,0.35) -- (\xValII,0.7)
-- (\xValIII,0.5);

\draw[thick,blue]
(\xValIII,0.5) -- (\xValIII,1.0)
-- (\xValIV,0);


\draw[thick,dashed,purple!80!black]
(0,0)
-- (\xValI,0)
-- (\xValII,0.35)
-- (\xValIII,0.5)
-- (\xValIV,0);


\begin{scope}[shift={(5.1,1.35)},scale=0.8]

\node[inner sep=3pt,draw,,minimum height=0.9cm,minimum width=2.05cm] (L) at (1,-0.25) {};

\draw[thick,blue] (0,-0.05) -- (0.6,-0.05)
    node[right] {\small $Q_k^\mathrm{buff}(t_q)$};

\draw[thick,dashed,purple!80!black] (0,-0.5) -- (0.6,-0.5)
    node[right] {\small $Q_k(t_q)$};

\end{scope}
\end{tikzpicture}
\caption{Illustration of the transformation from the physical queue $Q_k^\mathrm{buff}(t_q)$ to the virtual queue $Q_k(t_q)$. The colored intervals represent different operating regimes of the system, corresponding to varying relationships between the service rate $R_k$ and the desired target $\alpha^\mathsf{des}$. The virtual queue increases when the target $\alpha^\mathsf{des}$ is not met and decreases once the service rate exceeds $\alpha^\mathsf{des}$.}
\label{fig:QVirt}
\end{figure}
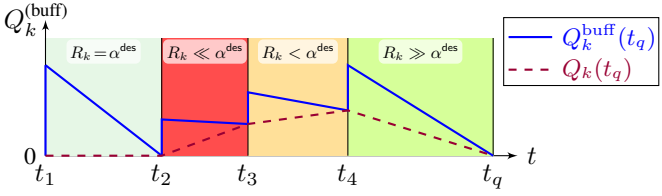

To isolate the effect of disruptions from the randomness of nominal traffic arrivals, we construct a virtual queue that evolves based on the long-term traffic statistics rather than the instantaneous packet arrivals. Accordingly, the random arrival process is replaced by its expected aggregate bit rate. This leads to the definition
\begin{align}
\alpha_k^\mathsf{des}
=
\xi_k \bar{M}_k^p,
\label{eq:alpha_k}
\end{align}
which corresponds to the nominal average number of arriving bits per second for user $k$. Hence, over a scheduling interval of duration $T_q$, the virtual queue is driven by a deterministic arrival of $T_q \alpha_k^\mathsf{des}$ bits. By substituting this expression into (\ref{eq:Q_base})
we can construct virtual queues as
\begin{align}\label{eq:virtQueue}
Q_k(t_{q+1}) = [Q_k(t_q) \underbrace{- \overbrace{{T_q} R_k(t_q)}^\text{departing bits} + \overbrace{T_q \alpha_k^\mathsf{des}}^{\text{arriving bits}}}_{\Delta Q_k(t_q)}]^+,
\end{align}
which now represent the effective backlog of data that has not been successfully transmitted and must be compensated in future transmissions. As illustrated in Fig.~\ref{fig:QVirt}, the virtual queue increases only when the system experiences a deviation from the desired operating condition, i.e., when the target \ac{QoS} $\alpha_k^\mathsf{des}$ is not met within the current time slot. This deviation may also arise due to operation in very short blocklength regimes as shown in Fig.~\ref{fig:flexTime}, where reduced transmission time leads to higher rate penalties and thus a higher likelihood of QoS violations.

In fact, the transformation into the virtual-queue domain can be interpreted as a temporal accounting mechanism for performance deviations, where deficits in service are accumulated over time and subsequently balanced within future transmissions. More generally, depending on the choice of the quantity of interest, this framework allows the virtual queue to be redefined to capture other forms of performance loss, thereby providing a flexible and unified representation of deviations from desired system objectives. For instance, when considering latency constraints, the virtual queue can represent accumulated delay violations, where unmet delay targets are carried over and must be compensated through faster service in subsequent intervals.

\subsection{Resilience Phases}
The behavior illustrated in Fig.~\ref{fig:QVirt} suggests that the evolution of the virtual queue is fundamentally governed by the balance between the achieved service rate and the desired \ac{QoS} target. To quantify this balance and characterize the resilience of the system with respect to virtual queue stability, we define the \emph{stability ratio} as the ratio of the instantaneous departure rate to the instantaneous arrival rate:
\begin{align}\label{eq:chi}
\chi_k = \frac{{R}_k}{{\alpha}_k^\mathsf{des}}.
\end{align}

\subsubsection{Normal Operation}
Based on (\ref{Q:steady}), the normal operation, or steady state, is achieved when the queues are stable, meaning that each queue's  departure rate is about equal to the arrival rate, i.e., $\chi_k~ \gtrsim~1$. This operating regime corresponds to the first interval in Fig.~\ref{fig:QVirt}, where $R_k=\alpha_k^\mathsf{des}$ and the virtual queue remains unchanged.
\subsubsection{Absorption}
When this balance is disturbed for a given user, e.g., due to traffic surges or channel degradation, the departure rate can no longer match the desired rate $\alpha_k^\mathsf{des}$, resulting in a stability ratio below one, i.e., $\chi_k < 1$. This behavior is illustrated by the second and third intervals in Fig.~\ref{fig:QVirt}, where $R_k<\alpha_k^\mathsf{des}$ and the virtual queue accumulates the resulting service deficit. In this regime, the queue begins to build up, with the growth rate determined by the deviation $1-\chi_k$. If this imbalance persists, the buffer may eventually overflow, leading to packet loss. The system remains in the absorption phase as long as the virtual queue continues to grow. Only once it stabilizes, it transitions into the adaptation phase, where resources are adjusted to restore nominal operating conditions.

\subsubsection{Adaptation}
Once the stability ratio $\chi_k$ returns to its nominal regime, i.e., $\chi_k \gtrsim 1$, the queue stops growing and the system enters the adaptation phase. This transition is illustrated by the final interval in Fig.~\ref{fig:QVirt}, where $R_k \gg \alpha_k^\mathsf{des}$ and the virtual queue begins to decrease. In this phase, the objective shifts from preventing further queue growth to actively reducing the accumulated backlog.

In this regime, the service rate exceeds the effective arrival rate, i.e., $\chi_k > 1$, which enables a compensation effect that gradually drains previously accumulated queue deficits and restores the queue to its pre-disruption state. Consequently, while the absorption phase focuses on preventing overflow under degraded conditions, the adaptation phase targets backlog recovery and system normalization. These fundamentally different objectives require distinct resource allocation strategies: the former prioritizes stability preservation, whereas the latter exploits additional degrees of freedom to accelerate queue depletion. This is particularly relevant in the finite blocklength regime, where transmission parameters can be tuned to enhance recovery speed, as detailed in the following sections.

\subsection{Resilience Metric} \label{sec:ResMetric}
To quantify the trade-offs that recovery under \ac{FBL} constraints imposes, we consider a resilience metric that captures a system’s capacity to absorb performance degradation, adapt to disturbances, and restore queue stability within an acceptable timeframe.
Following the guidelines in~\cite{resiliencemetric,RobertRes}, we define a \textit{Queue State Index (QSI)} that takes the value \(1\) during normal operation and \(0\) when recovery has failed. To this end, we formulate \vspace{-0.1cm}
\begin{align}\label{func_fun}
\mathcal{F}(t_q) = \prod_{k=1}^{K} \Biggl[\frac{ Q_k^\mathsf{max} - Q_k(t_q) }{Q_k^\mathsf{\max} - Q_k^{\mathsf{des}}} \Biggr]_0^1,
\end{align}
where \(Q_k^{\max}\) is user $k$'s maximum buffer size, \(Q_k^{\mathsf{des}}\) is user $k$'s desired queue size and $[x]_0^1 = \min\{\max\{x,0\},1\}$.
If \(Q_k(t_q) > Q_k^{\max}\) for any user, the corresponding factor in (\ref{func_fun}) becomes zero. Conversely, if \(Q_k(t_q) < Q_k^\mathsf{des}\), the value of the user’s QSI is saturated at 1. The resulting QSI for the system is defined as the product of every user's $\mathcal{F}(t_q)$. This is essential because the loss of data for even a single user can propagate into system-wide or catastrophic failures, a risk that is especially critical in \ac{URLLC} applications.
\subsubsection{Absorption Metric}
Regarding the temporal component, we let \(t_{\mathsf{out}}\) denote the time of the outage event, and \(t_{\mathsf{abs}}\) the time at which the failure has been absorbed. That is, when all queues, i.e., transmit queues, have stabilized and can be defined as
\begin{align}\label{t_abs}
t_{\mathsf{abs}} \;=\; \min \big\{ t_q > t_{\mathsf{out}} \;:\;\chi_k \geq 1, \; \forall k \in \mathcal{K} \big\},
\end{align}
where $q_{\mathsf{out}} := \min \{q : t_q \ge t_{\mathsf{out}}\}$,
denotes the first discrete-time index corresponding to the outage event time $t_{\mathsf{out}}$. Similarly, the index $q_{\mathsf{abs}} := \min \{q : t_q \ge t_{\mathsf{abs}}\}$ maps the absorption time $t_{\mathsf{abs}}$ to its corresponding discrete-time representation.
Within the absorption phase, the primary objective is to stabilize the queue states of all users as quickly as possible, thereby preventing any queue overflow.

To consider both, the time needed for successful absorption and the level of maintained functionality, we define an intuitive absorption metric using a time-weighted version of the QSI:
\begin{align}\label{eq:r_abs}
r_{\mathrm{abs}} =
\frac{
\sum_{q = q_{\mathsf{out}}}^{q_{\mathsf{abs}} - 1}
\mathcal{F}(t_q)\, T_q
}{
\sum_{q = q_{\mathsf{out}}}^{q_{\mathsf{abs}} - 1}
T_q
}
\in [0,1].
\end{align}
Here, the nominator spans the area under the resilience curve, while the denominator spans the whole area that would be achieved during normal operation. Due to the time-weighting, this ratio considers not only the final stabilized functionality at $t_\mathsf{abs}$ but also the shape of the degradation during the absorption phase: a rapid initial drop and a long convergence time both increase the lost area under the curve, meaning the system effectively operated at a lower average performance throughout the absorption phase. Consequently, early improvements contribute more to the average value, meaning systems that restore functionality quickly will achieve a higher score than those that recover late, even if both reach the same final state.
\subsubsection{Adaption Metric}
After reaching the final state in the absorption phase, the system enters the {adaptation phase}, which begins at the absorption time $t_{\mathrm{abs}}$ and terminates at the time-to-recovery $t_{\mathsf{ada}}$.  Intuitively, $t_{\mathsf{ada}}$ is the earliest instant after $t_{\mathsf{abs}}$ at which the queues have returned to (or fallen below) their pre-disruption operating point and the stability condition is reestablished.
To this end, we define the end of the adaption phase as the time-to-recovery as
\begin{equation}\label{t_rec}
t_{\mathsf{ada}} =
\min\big\{ t > t_{\mathrm{abs}} \;:\;
Q_k(t) \le Q_k^{\mathrm{des}},\;
\chi_k(t) \ge 1,\ \forall k \in \mathcal{K}
\big\},
\end{equation}
where the corresponding discrete-time index is defined as\vspace{-0.1cm}
\begin{align}
q_{\mathsf{ada}} := \min \{q : t_q \ge t_{\mathsf{ada}}\}.
\end{align}

In order to satisfy the condition $\chi_k(t)\geq 1$, the system’s departure rates must exceed the arrival rates. Achieving this can be challenging, especially during the disruption, and it may take a significant amount of time. To provide a more insightful measure of resilience, we divide the adaptation phase into two complementary metrics that separately capture recovery performance and the time required to recover.

The adaptation performance metric should measure how effectively the system recovers relative to the losses incurred during the absorption phase. To this end, we define it as follows:
\begin{align}\label{eq:r_ada}
r_{\mathsf{ada}} =
\frac{
\sum_{q = q_{\mathsf{abs}}}^{q_{\mathsf{ada}} - 1}
\big(\mathcal{F}(t_{q}) - \mathcal{F}(t_{\mathsf{abs}})\big)\, T_{q}
}{
\big(1 - \mathcal{F}(t_{\mathsf{abs}})\big)
\sum_{q = q_{\mathsf{abs}}}^{q_{\mathsf{ada}} - 1}
T_q
}.
\end{align}
Here, $r_{\mathsf{ada}} \in [0,1]$ quantifies the normalized recovery of the system relative to the performance level at the end of the absorption phase. A value of $r_{\mathsf{ada}} = 0$ indicates that no recovery is achieved after absorption, i.e., the system remains at the degraded performance level $\mathcal{F}(t_{\mathsf{abs}})$. In contrast, $r_{\mathsf{ada}} = 1$ indicates full recovery, meaning that all queues satisfy the desired operating condition $Q_k(t_q) \le Q_k^{\mathrm{des}}$ for all $k \in \mathcal{K}$.

\subsubsection{Relative Recovery Index (RRI)}
While $r_{\mathrm{abs}}$ and $r_{\mathsf{ada}}$ incorporate temporal effects through time-weighted performance evolution within each phase, they do not explicitly compare the characteristic timescales of different phases. This motivates the introduction of a relative recovery metric $r_{\mathsf{rec}}$, which normalizes the adaptation duration with respect to the absorption phase.
We define the Relative Recovery Index (RRI) as 
\begin{align}\label{eq:r_rec}
    r_{\mathsf{rec}} = 2^{ -\left(\frac{T_{\mathsf{ada}}}{T_{\mathsf{abs}}}-1\right)},
\end{align}
where $T_{\mathsf{ada}} = t_{\mathsf{rec}} - t_{\mathsf{abs}}$ denotes the duration of the adaptation phase, and $T_{\mathsf{abs}} = t_{\mathsf{abs}} - t_{\mathsf{out}}$ denotes the duration of the absorption phase.
The relative formulation is motivated by the fact that temporal effects are already captured in $r_{\mathrm{abs}}$ and $r_{\mathsf{ada}}$, which account for performance evolution over time. In contrast, $r_{\mathsf{rec}}$ isolates the temporal efficiency of recovery by normalizing it with respect to the absorption timescale. This avoids penalizing longer disturbances in absolute terms while still rewarding faster recovery relative to the outage duration.

\subsubsection{Unified Resilience Metric}
Subsequently, the three introduced metrics are combined into a unified resilience metric:
\begin{align}\label{eq:r}
    r = \lambda_1 r_\mathsf{abs} + \lambda_2 r_\mathsf{ada} + \lambda_3 r_\mathsf{rec} \in [0,1],
\end{align}
where $\lambda_i \geq 0$ and $\sum_{i=1}^{3} \lambda_i = 1$, reflecting the network’s prioritization of robustness, adaption quality, and temporal efficiency. Importantly, each of the individual metrics is defined in a way that provides direct, interpretable insight. This makes it easy to evaluate the system’s resilience behavior and to identify which aspects of the applied resilience methods are strong and which require improvement. This intuitive interpretation is further demonstrated later in the numerical results section.

\section{Problem Formulation}
In this section, we formulate a unified framework that optimizes the resilience metric from physical layer perspective. To this end, we split the recovery procedure in two optimization problems that tackle the diverse and distinct properties of the absorption and adaption phase, respectively. To do so, we transform the objective into a Lyapunov function because the resilience metrics's evaluation is a long-term objective within both phases. The evaluation of its stability results in a per time slot objective to guide the optimization.

\subsection{Lyapunov-Based Transformation and Stability Indicators}
The core of the resilience metric is based on the QSI $\mathcal{F}(t_q)$, which is dependent on the buffer states of each user. In order to efficiently guide this function into the desired states defined by $t_\mathsf{abs}$ in (\ref{t_abs}) and $t_\mathsf{rec}$ in (\ref{t_rec}), which are long-term goals, we reformulate the QSI into a Lyapunov function, which allows us to determine suitable per time-slot control rules for asymptotic stability. In addition, our problem poses hard bounds which are met when the transmit buffers overflow. To achieve stability with these constraints, we first define a suitable log-barrier Lyapunov function.

In the context of resilience, we are particularly interested in fast time-slot–by–time-slot control, as this allows the system to respond promptly and recover from outages. To achieve stability under these considerations, we define a suitable log-barrier Lyapunov function $V$ as
\begin{align}\label{eq:V_k}
  V(\mat{Q}(t_q)) = - \sum_{k\in\mathcal{K}} \log\left(\frac{Q_k^\mathsf{max}-Q_k(t_q)}{Q_k^\mathsf{max}-Q_k^\mathsf{des}}\right),
\end{align}
where $\mat{Q}(t_q)= [Q_1(t_q), \dots, Q_K(t_q)]$ is a vector containing the users' queues in time slot $q$.
For the defined Lyapunov function $V(\mat{Q}(t_q))$, the per time slot increment can be written as
\begin{align}\label{eq:Lyap_slot}
  \Delta V(t_q) = V(\mat{Q}(t_q)+\Delta \mat{Q}(t_q)) - V(\mat{Q}(t_q)),
\end{align}
where $\Delta \mat{Q}(t_q) = [\Delta {Q_1}(t_q), \dots, \Delta {Q_K}(t_q)]$ is a vector consisting of the changes in each user's queue as defined in (\ref{eq:Q_base}). By substituting $\Delta\mat{Q}(t_q)$ into (\ref{eq:Lyap_slot}) and simplifying, we obtain the Lyapunov drift term as
\begin{align}\label{deltaV}
  \Delta V(t_q)  &= -\sum_{k\in \mathcal{K}} \log\left(1- \frac{\Delta Q_k (t_q)}{Q_k^\mathsf{max}-Q_{k}(t_q)} \right)\\
 \Delta V(t_q)  &= - \sum_{k\in \mathcal{K}} \log(1-\delta_{k}^q),
\end{align}
where
\begin{align}
  \delta_{k}^q & = \frac{-T_q R_{k}(t_q) + T_q\alpha_k^\mathsf{des}}{Q_k^\mathsf{max}-Q_{k}(t_q)}
\end{align}
is a per-user increment term that can be understood as the normalized growth factor of user $k$'s queue in time slot $q$.
%
These expressions also clarify how the log-barrier ensures stability. Since
$-\log(1 - \delta_k)$ grows rapidly as $\delta_k \rightarrow 1$, the Lyapunov
increment becomes large whenever a queue approaches its limit, signaling
instability and resulting in a critical failure state. For $0 < \delta_k < 1$, the term is positive but bounded,
indicating moderate queue growth and the closer $\delta_k$ is to zero, the smaller
the drift and the more stable the system behaves. If $\delta_k \leq 0$, the
argument of the logarithm exceeds one, yielding a negative drift that pushes
the system back toward the desired operating region. This makes $\delta_k$ an
effective indicator for guiding resilience-aware optimization.

\subsection{Absorption Phase}
Within the absorption phase, we are interested in stabilizing the queues as fast as possible after a disruption causes the rates to drop below the required \ac{QoS}. At the same time, the system must react in a way that prevents any user from entering the failure state, i.e., $Q_k(t_q) \geq Q_k^\mathsf{max}$. Based on the formulated metric $r_\mathsf{abs}$, the main objective becomes the minimization of the absorption duration $t_\mathsf{abs}$ while also maximizing the QSI $\mathcal{F}(t_q)$ over all slots. To enable per-timeslot control without requiring knowledge of future states, we adopt the normalized growth factor $\delta_k^q$ as a Lyapunov-based indicator for the stability of the system. As such, we can ensure stability by formulating the optimization problem as
\begin{align}\label{P1}\tag{P1}
&\hspace{-2cm}\underset{\eta_q, \vect{r}^q ,\vect{v}^q ,\{\vect{w}_k^q \}_{k\in\mathcal{K}},\,}{\min} \quad \eta_q\\
\text{s.t.}\quad &(\ref{eq:powConst}),\nonumber \\
& \frac{-\eta_q (r_{k}(t_q) - \alpha_k^\mathsf{des})}{B(Q_k^\mathsf{max}-Q_{k}(t_q))} \leq 0, \quad \forall k \in \mathcal{K},  \label{eq:stability_constr}\\
& r_{k}^q \leq C_k^q -  B\Omega \, \sqrt{{V_\mathsf{disp}(\Gamma_{k}^q)}/{\eta_q}} , \label{eq:r_expl_eta}\\
& |v_{m}^q| = 1, \quad \forall m \in \{1,\dots,M\},\label{psConst}  \\
& \eta_{\min} \leq \eta_q \leq \eta_{\max}, \label{etaConst}
\end{align}
where $\vect{r}^q = [r_1^q, r_2^q, \dots, r_K^q]^T$ represents the stacked rate vector of time slot $q$. The unit-modulus constraints in (\ref{psConst}) enforce the phase shift conditions $0\leq \theta_m^q \leq 2\pi, \forall m \in \{1,\dots,M\}$ and $\eta_{\min}$ and $\eta_{\max}$ are the respective minimum and maximum blocklengths, which depend on the capabilities or constraints of the utilized system. However, the overall problem is non-convex due to the coupling of the variables. Further, the rate expression in~(\ref{eq:r_expl_eta}) is non-convex in both the beamforming vector and the RIS phase shifts, and the unit-modulus constraint in (\ref{psConst}) induces a non-convex feasible set. Furthermore, the blocklength $\eta_q$ enters the rate through the finite-blocklength penalty term and simultaneously determines the duration of the current time slot, thereby introducing additional non-convexity and making the optimization problem non-trivial. We address these interdependencies by utilizing state-of-the-art techniques such as \ac{AO} and \acp{SCA} \cite{SynBenefits,accelRec}.

Under these considerations, we reformulate problem (\ref{P1}) into an equivalent but more tractable form, enabling efficient application of \ac{AO} and \ac{SCA} under \ac{FBL} constraints. To this end, we decouple the \ac{SINR} and channel dispersion constraints by introduce the slack variables $\vect{\gamma}^q=[\gamma_1^q ,\dots ,\gamma_K^q]$ and $\vect{u}^q=[u_1^q ,\dots ,u_K^q]$ as
\begin{align}
\label{P1.1}
\tag{P1.1}
\underset{\vect{\gamma},\,\vect{u},\,\eta, \vect{r} ,\vect{v} ,\{\vect{w}_k \}_{k\in\mathcal{K}},\,}{\min}
 \eta,\\
\quad  \text{s.t.} \quad  \text{(\ref{eq:powConst})}, \text{(\ref{eq:stability_constr})}, \text{(\ref{psConst})},& \text{(\ref{etaConst})} \nonumber \\
 r_{k} \leq B \big( \log_2(1+\gamma_{k})& - \frac{\Omega}{\sqrt{\eta}} u_k \big), \, &&\forall k \in \mathcal{K}, \label{P2rate}\\
\gamma_k  \,\,\ &\leq \Gamma_k , \, &&\forall k \in \mathcal{K}, \label{P2SINR}\\
u_k \,\,\  &\geq \sqrt{V_\mathsf{disp}(\gamma_k)}  &&\forall k \in \mathcal{K},\label{gradVP1}\\
		\:\,\:\,\vect{\gamma} \:\,&\geq 0 , \,\, \vect{u} \:\,\geq 0,\label{P2t1}
\end{align}
where we omit the index $q$ and dependency on $t_q$ for clarity, since the optimization problem is solved independently in each time slot. Furthermore, (\ref{P2t1}) ensures that all entries in $\vect{q}$, and similarly in $\vect{u}$, are nonnegative. Nevertheless, due to the coupling between $\eta$ and $u_k$ in (\ref{P2rate}), as well as the coupling between $\vect{w}_k$ and $\vect{v}$ in (\ref{P2SINR}), the problem remains non-convex. To further address these problems, we decompose the problem into three subproblems, in which we optimize the beamforming vectors, the phase shifts, and the blocklengths independently.

\subsubsection{Beamforming Optimization}
Keeping $\vect{v}$ and $\eta_q$ fixed, the problem reduces to verifying whether the stability conditions in \eqref{eq:stability_constr} can be satisfied, i.e., to a feasibility detection problem. However, pure feasibility problems provide no information about the degree of violation when the constraints are not met and may therefore lead to slow or unstable convergence.

To enable a directed and rapid approach toward the stability region while still aiming to minimize the blocklength, we instead minimize the maximum stability violation. Specifically, we replace \eqref{eq:stability_constr} by the min–max formulation
\begin{align}\label{eq:minmax}
\underset{\vect{\gamma},\,\vect{u},\, \vect{r}  ,\{\vect{w}_k \}_{k\in\mathcal{K}},\,}{\min} \underset{k}{\max} \quad \delta_k,
\end{align}
which directly targets the worst-case constraint violation. This reformulation ensures that, in each iteration, optimization effort is concentrated on the most critical user, thereby driving the system toward feasibility as quickly as possible. Once the optimal value becomes non-positive, the stability conditions are satisfied for all users.
By introducing an auxiliary variable $\zeta$, the min–max objective in \eqref{eq:minmax} can be equivalently reformulated in epigraph form as
\begin{align}
\label{P1.2}
\tag{P1.2}
\underset{\zeta,\vect{\gamma},\,\vect{u},\,\vect{r} ,\{\vect{w}_k \}_{k\in\mathcal{K}},\,}{\min}
 \zeta\\
\quad  \text{s.t.} \quad  \text{(\ref{eq:powConst})},& \text{(\ref{P2rate})},\text{(\ref{P2SINR})}, \text{(\ref{gradVP1})},\text{(\ref{P2t1})}, \nonumber \\
&\frac{-\eta (r_{k} - \alpha_k^\mathsf{des})}{B(Q_k^\mathsf{max}-Q_{k})} \leq \zeta. \label{eq:epigraph}
\end{align}
Following \cite{RIS_RES,accelRec}, the first-order Taylor approximation around the point 	$(\tilde{\vect{w}},\tilde{\vect{\gamma}})$ can be applied to the \ac{SINR} constraints. The resulting convex approximation of (\ref{P2SINR}) can be written as
	\begin{align}\label{SINR_convex1}
	\sum_{i\in\mathcal{K}\setminus\{k\}}|(\RisChan{k})^H \vect{w}_i|^2 + \sigma^2 +
			\frac{|(\RisChan{k})^H \tilde{\vect{w}}_k|^2}{(\tilde{\gamma}_k)^2}\gamma_k \nonumber \\ - \frac{2 \text{Re} \{\tilde{\vect{w}}_k^H(\RisChan{k})(\RisChan{k})^H\vect{w}_k \}}{\tilde{\gamma_k}} \leq 0, \forall k \in \mathcal{K}.
\end{align}
Regarding the term in (\ref{gradVP1}), we can derive the first-order Taylor approximation around the point $\tilde{\gamma}_k$ \cite{Cleckx_approx} as
\begin{align}\label{eq:ChanDispApp}
 \sqrt{V(\gamma_k)} \leq& \sqrt{1 - (1+\tilde{\gamma}_k)^{-2}} + \big( 1+\tilde{\gamma}_k  \big)^{-3}\nonumber\\ &(1 - (1+(\tilde{\gamma}_k)^{-2}))^{-\frac{1}{2}} (\gamma_k - \tilde{\gamma}_k)  \triangleq  U_k(\gamma_k).
\end{align}
Using these convexifications, the overall beamforming optimization problem can be formulated as
\begin{align}\label{P1.3}\tag{P1.3}
\underset{\zeta,\vect{\gamma},\,\vect{u},\,\vect{r} ,\{\vect{w}_k \}_{k\in\mathcal{K}},\,}{\min}
 \zeta\\
\quad  \text{s.t.} \quad  \text{(\ref{eq:powConst})},& \text{(\ref{P2rate})}, \text{(\ref{P2t1})}, \eqref{eq:epigraph}, \eqref{SINR_convex1}, \eqref{eq:ChanDispApp}. \nonumber
\end{align}
In this state, problem \eqref{P1.3} can be solved iteratively by using the \ac{SCA} method.
More specifically, we denote $\mathbf{\Lambda}_z^w = \left[\{\vect{w}_{k,z}^T\}_{k \in \mathcal{K}}, \boldsymbol{\kappa}_z^T \right]^T$
as a vector stacking the optimization variables of the beamforming design problem at iteration $z$, where $\vect{\kappa}_z = [\vect{r}_z^T,\vect{\gamma}_z^T,\vect{u}_z]^T$. Similarly $\hat{\mat{\Lambda}}_z^w  = [\{\hat{\vect{w}}_{k,z}^T\}_{k \in \mathcal{K}} ,\hat{\vect{\kappa}}_z^T ]^T$ and $\tilde{\mat{\Lambda}}_z^w  = [\{\tilde{\vect{w}}_{k,z}^T\}_{k \in \mathcal{K}} ,\tilde{\vect{\kappa}}_z^T ]^T$ define the optimal solutions and the point, around which the approximations are computed, respectively. Based on these expressions, given a point $\tilde{\mat{\Lambda}}_{z}^w $, we can obtain an optimal solution $\hat{\mat{\Lambda}}_z^w $ by solving problem \eqref{P1.3}.
\subsubsection{Phase Shift Optimization}
During the design of the phase shifts at the \ac{RIS}, the beamforming vectors are held fixed, following the alternating optimization strategy. To maintain consistency with the problem structure in (\ref{P1.2}), we define $(\RisChan{i})^H \vect{w}_k$ $=$ $\tilde{h}_{i,k}^* + \tilde{\mat{G}}_{i,k}^* \vect{v}^* = \beta_{i,k}$, where $\tilde{h}_{i,k}^* = \vect{w}_k^H \vect{h}_i$ and $\tilde{\mat{G}}_{i,k}^* = \vect{w}_k^H \mat{G}_i$, where $(\cdot)^*$
denotes the complex conjugate. With these definitions, the \ac{SINR} constraints in (\ref{P2SINR}) can be approximated using a similar procedure as applied in the beamforming design. Following the approach in \cite{RIS_RES,accelRec}, the first-order Taylor approximation around the point $(\tilde{\vect{v}}, \tilde{\vect{q}})$ can be calculated as
\begin{align}\label{apprxV}
\sum_{i\in\mathcal{K}\setminus\{k\}}&|\beta_{k,i}|^2 + \sigma^2 -\frac{|{\beta_{k,k}}|^2}{\tilde{\gamma}_k }-\frac{2}{\tilde{\gamma}_k} \Re\left\{ \beta^*  \vect{G} \left( \vect{v} - \tilde{\vect{v}} \right) \right\} \nonumber\\
&\qquad \qquad \quad +\frac{|\beta_{k,k}|^2}{\tilde{\gamma}_k^2} \left( \gamma_k - \tilde{\gamma}_k \right) , \quad \forall k \in \mathcal{K}
\end{align}
To handle the unit-modulus constraint in (\ref{psConst}), we adopt the penalty method proposed in \cite{RIS_RES}. Specifically, the term $\sum_{m=1}^{M}(|v_m|^2 -1)$ is approximated using a weighted first-order Taylor expansion around a given point $\tilde{\vect{v}}$. The resulting expression,
\[
\Phi = \alpha_{v}\sum_{m=1}^{M}\text{Re}\left\{2\tilde{v}_m^*v_m - |\tilde{v}_m|^2\right\},
\]
is then incorporated into the objective function as a penalty, where $\alpha_{v} \gg 1$ is a large constant controlling the strength of the penalization.

At this point, the approximated optimization problem for the phase-shift design can be formulated as
\begin{align}\label{P2}\tag{P2}
	\underset{\zeta, \vect{v},\vect{r},\vect{\gamma},\vect{u}}{\min}  \,\,\, \,\,& \zeta + \Phi \\
	\text{s.t.} \quad &  \text{(\ref{P2rate})},\text{(\ref{P2t1})},\eqref{eq:epigraph}, \text{(\ref{eq:ChanDispApp})}, \text{(\ref{apprxV})},
 \nonumber
\end{align}
where the epigraph reformulation is employed in a manner analogous to \eqref{P1.2}. Due to the similarity of the problem formulation and the utilization of the same \ac{SCA} framework, problem (\ref{P2}) can be solved by defining $\mat{\Lambda}_z^v = [\vect{v}_z^T, \vect{\kappa}_z^T]^T$ and following the same iterative procedure as for solving problem (\ref{P2}).
\subsubsection{Blocklength/Transmit Duration Optimization}
During the optimization of the blocklength $\eta$ for the current timeslot $t_q$, the other optimization variables are kept fixed due to the \ac{AO} approach. Since the objective variable and the optimization variable coincide for this case,
the problem can also be can be equivalently reformulated to minimize the maximum stability violation similar to \eqref{eq:minmax}. As such Problem (\ref{P1}) can be formulated as
\begin{align}\tag{P3}\label{eq:P3}
\underset{\eta}{\min}&\quad \underset{k}{\max} \quad \delta_k,\\
\text{s.t.} & \quad  \eqref{eq:stability_constr}, \text{(\ref{eq:r_expl_eta}), (\ref{etaConst})}.\nonumber
\end{align}
By substituting the rate constraint \eqref{eq:r_expl_eta} into the stability constraint \eqref{eq:stability_constr}, and exploiting that the rate constraint is active at the optimum,
the stability condition can be written as
\begin{align}\label{eq:etaMinConst}
\bar{\delta}_k \triangleq
\frac{-\eta \left(
C_k
- B\Omega \sqrt{\frac{V_\mathsf{disp}(\Gamma_k)}{\eta}}
- \alpha_k^\mathsf{des}
\right)}
{B\big(Q_k^\mathsf{max}-Q_k\big)}
\le 0,
\quad \forall k \in \mathcal{K}.
\end{align}
Hence, the original problem is equivalently reformulated as the
following min--max optimization:
\begin{align}\label{eq:minmax_delta}
\min_{\eta_{\min} \le \eta \le \eta_{\max}}
\;\max_{k\in\mathcal{K}} \; \bar{\delta}_k.
\end{align}
Since $\bar{\delta}_k$ is continuously differentiable with respect to
$\eta$, the optimal solution of \eqref{eq:minmax_delta} is obtained
by solving the first-order optimality condition of the active (worst-case)
user, i.e., by equating $\frac{\partial \bar{\delta}_k}{\partial \eta}=0$
for the maximizing index $k$. The resulting stationary point is then
projected onto the feasible interval $[\eta_{\min},\eta_{\max}]$ and yields the following analytical solution:
\begin{align}
\eta^* = \left[ \max_{k\in\mathcal{K}} \frac{(B \Omega)^2 V_\mathsf{disp}(\Gamma_k^q)}{4 (C_k^q - \alpha_k^\mathsf{des})^2} \right]_{\eta_\mathsf{min}}^{\eta_\mathsf{max}}.
\end{align}
The detailed derivation of the resulting analytical solution is provided in Appendix \ref{app:A}.

\subsection{Adaption Phase}
After stabilizing the queues in the absorption phase, the virtual queues no longer grow, ensuring system stability. The recovery constraints in \eqref{t_rec} differ only by
\begin{align}
Q_k(t_q) \le Q_k^{\mathrm{des}}, \quad \forall k \in \mathcal{K}, \label{eq:Q_const}
\end{align}
which requires compensating the service deficit accumulated during the blockage, where only users affected by \eqref{eq:Q_const} need to be explicitly considered.

However, the quality of adaptation in this phase is evaluated differently, namely through the performance metric $r_\mathsf{ada}$. Since the QSI $\mathcal{F}(t_q)$ embedded in $r_\mathsf{ada}$ assigns a higher value to multiple small deviations than to a single dominant deviation, the resource reallocation strategy in the adaptation phase differs from that in the absorption phase.

Moreover, due to the initial resource loss caused by the failure and the additional resource consumption required to absorb the deficit, mobilizing sufficient resources to not only match, but also surpass the nominal \ac{QoS} level $\alpha_k^\mathsf{des}$ becomes highly challenging. Therefore, we propose a reformulated optimization approach that more appropriately captures the objectives of the adaptation phase, while freeing up some resources to enable quick adaption.
More specifically, we are interested in minimizing the overall backlog in the virtual queues of each user $k \in \mathcal{K}$ at time $t_q$ given as
\begin{equation}
\tilde Q_k(t_q) = Q_k(t_q) - Q_k^\mathrm{des}.
\end{equation}
We define the standard Lyapunov function over the residual backlogs as
\begin{equation}
V^\mathsf{ada}(t_q) = \frac{1}{2} \sum_{k \in \mathcal{K}} \tilde Q_k(t_q)^2.
\end{equation}
Considering the queue dynamics in \eqref{eq:virtQueue}, the Lyapunov drift over one slot becomes
\begin{align}
\Delta V^\mathsf{ada} =& V^\mathsf{ada}(t_{q+1}) -  V^\mathsf{ada}(t_q)
= \sum_{k \in \mathcal{K}} \tilde Q_k(t_q) \big( \alpha_k^\mathsf{des} - R_k \big) T_q
\end{align}
This leads to the following optimization problem:
\begin{align}\label{P4.1}\tag{P4.1}
&\hspace{-1cm}\underset{\eta_q, \vect{r}^q ,\vect{v}^q ,\{\vect{w}_k^q \}_{k\in\mathcal{K}}}{\min} \ \sum_{k \in \mathcal{K}} \tilde Q_k(t_q) \big( \alpha_k^\mathsf{des} - r_k \big) T_q \nonumber\\
\text{s.t.}\quad &(\ref{eq:powConst}), \eqref{eq:r_expl_eta} - \eqref{etaConst},\\
&\frac{-\eta (r_{k} - \alpha_k^\mathsf{des})}{B(Q_k^\mathsf{max}-Q_{k})} \leq \zeta^\mathsf{Thr} \label{eq:stabThresh}
\end{align}
which ensures that all residual backlogs are reduced proportionally to their current size, driving the system toward an equilibrium in which each queue approaches its desired value $Q_k^\mathrm{des}$ simultaneously. Furthermore, the stability constraints in \eqref{eq:stability_constr} are replaced, as the solution obtained in the absorption phase can be leveraged to ensure long-term stability once the recovery constraint in \eqref{eq:Q_const} is satisfied. To enhance flexibility under heterogeneous channel conditions, we adopt the relaxed stability constraint in \eqref{eq:stabThresh}. In particular, allowing the stability threshold to take values $0 < \zeta^\mathsf{Thr} \ll 1$ permits temporary deviations from strict queue stability, such that the queues of users that are currently below $Q_k^\mathsf{des}$ may increase. This controlled relaxation, however, enables a more efficient redistribution of resources toward users in critical conditions, i.e., those that are stabilized close to their buffer limit $Q_k^\mathsf{max}$.

\subsubsection{Beamforming Optimization}
Similarly to the assumptions in the absorption phase, the phase-shift vector $\vect{v}$ and blocklength $\eta$ are fixed during the beamforming optimization. Consequently, the beamforming problem within the adaption phase can be expressed similarly as
\begin{align}\label{P5}\tag{P5.1}
\underset{\vect{\gamma},\,\vect{u},\,\vect{r} ,\{\vect{w}_k \}_{k\in\mathcal{K}},\,}{\min}&
 \sum_{k \in \mathcal{K}} \tilde Q_k(t_q) \big( \alpha_k^\mathsf{des} - r_k \big) T_q\\
\quad  \text{s.t.} \quad  \text{(\ref{eq:powConst})},& \text{(\ref{P2rate})}, \text{(\ref{P2t1})}, \eqref{SINR_convex1}, \eqref{eq:ChanDispApp}, \eqref{eq:stabThresh}. \nonumber	
\end{align}

\subsubsection{Phase Shift Optimization}
For the phase shift optimization within the adaption phase, the problem can also be formulated with the same set of convex constraints as in the absorption phase. Using the same assumptions and definitions as for problem \eqref{P2}, we can write
\begin{align}\label{P6}\tag{P6}
\underset{\zeta, \vect{v},\vect{r},\vect{\gamma},\vect{u}}{\min}  \,\,\, \,\,& \zeta + \Phi \\
	\text{s.t.} \quad &  \text{(\ref{P2rate})},\text{(\ref{P2t1})},\eqref{eq:epigraph}, \text{(\ref{eq:ChanDispApp})}, \text{(\ref{apprxV})},\eqref{eq:stabThresh}.\nonumber
\end{align}
\subsubsection{Blocklength Optimization}
When optimizing the blocklength of the transmissions, we can reduce to problem to
\begin{align}\label{P7}\tag{P7}
\underset{\eta}{\min}&
 \sum_{k \in \mathcal{K}} \tilde Q_k(t_q) \big( \alpha_k^\mathsf{des} - r_k \big) \eta/B\\
\quad  \text{s.t.} \quad \eqref{eq:stabThresh}. \nonumber	
\end{align}
The optimal blocklength for the objective function $\eta_\text{obj}^\star$ is obtained in closed form as
\begin{align}\label{eq:nobj}
\eta_\text{obj}^\star &=
\left(
\frac{\Omega \sum_{k \in \mathcal{K}} \tilde Q_k u_k}
{-2 \sum_{k \in \mathcal{K}} \tilde Q_k (\alpha_k^\mathsf{des} - C_k)/B}
\right)^2,
\end{align}
while the relaxed per-user stability constraint yields
\begin{align}\label{eq:nthr}
\eta_k^\mathsf{Thr} &=
\biggl(
\frac{\vartheta_k - \sqrt{\vartheta_k^2 - 4 (C_k - \alpha_k^\mathsf{des}) \, \zeta^\mathsf{Thr} B (Q_k^\mathsf{max}-Q_k)}}
{2 (C_k - \alpha_k^\mathsf{des})}
\biggr)^2,
\end{align}\label{eq:etaOpt}
\noindent with $\vartheta_k = B \Omega u_k$.
The detailed derivation of the resulting analytical solution for $\eta_\text{obj}^\star$ and $\eta_k^\mathsf{Thr}$ is provided in Appendix \ref{app:B}. The final optimal blocklength is then determined as
\begin{align}\label{eq:nStar}
\eta^\star = \left[\min \Big( \eta_\text{obj}^\star, \; \min_{k \in \mathcal{K}} \eta_k^\mathsf{Thr} \Big)\right]^{\eta_\mathsf{max}}_{\eta_\mathsf{min}},
\end{align}
ensuring compliance with the stability and blocklength size constraints.

\subsection{Two-Phase Temporal Resilience Optimization}
In order to guarantee high performance within the context of the proposed resilience metric in \eqref{eq:r}, the objective during the optimization process is to absorb and if necessary adapt to blockages quickly. To this end, we utilize the following \ac{AO} algorithm, which starts with a feasible allocation of resources and triggers if blockages are detected. More precisely the algorithm first applies rate adaption by equalizing the expressions in \eqref{FBL_rate}, such that all users are still able to decode their symbols albeit at possibly lower rate. Afterwards, the absorption phase begins, in which the algorithm tries to stabilize the rate gap utilizing \eqref{P1}. If a feasible stable solution is found, the values are stored and the algorithm proceeds to the adaption phase, in which the accumulated overhead in the virtual queue is reduced to the nominal level by solving problem \eqref{P4.1}. After successful restoration of the queues, the stable solution from the absorption phase is applied and considered the new nominal operating state. Following each inner iteration of the absorption and adaptation phases, the algorithm also checks for any new outages. If such events are detected, the procedure is restarted.

\subsubsection{Complexity Analysis}
The computational complexity of the proposed resilience-oriented \ac{AO} framework, comprising the absorption and adaptation phases, is dominated by the beamforming and RIS phase-shift optimization subproblems, both solved via \ac{SCA}. In each phase, the beamforming problem involves $\mathcal{O}(KL)$ variables and $\mathcal{O}(K+N)$ constraints, resulting in a per-iteration complexity of $\mathcal{O}((KL)^3)$ using interior-point methods, while the RIS optimization problem scales with $\mathcal{O}(M+K)$ variables and has complexity $\mathcal{O}((M+K)^3)$. In contrast, the blocklength optimization admits a closed-form solution with linear complexity $\mathcal{O}(K)$ and is therefore negligible. Letting $I_{\mathrm{AO}}$, $I_w$, and $I_v$ denote the numbers of \ac{AO} and \ac{SCA} iterations, the overall complexity is $\mathcal{O}\big(I_{\mathrm{AO}}[I_w (KL)^3 + I_v (M+K)^3]\big)$. The proposed method avoids high-complexity semidefinite relaxation, ensuring scalability, with the dominant computational cost determined by either the antenna dimension $KL$ or the number of RIS elements $M$.
\subsubsection{Convergence Analysis}
The proposed algorithm is based on an \ac{AO} framework combined with \ac{SCA}. Each subproblem is solved optimally within its convex surrogate, which is tight at the current iterate and satisfies standard first-order consistency conditions. As a result, the objective function is monotonically non-increasing and bounded from below due to the imposed power and blocklength constraints. Therefore, the proposed two-phase algorithm is guaranteed to converge to a stationary (KKT) point of the original non-convex problem.

\begin{algorithm}[t]
	\caption{Two-Phase Temporal Resilience Optimization}
	\label{alg:alt_opt}
\small
	\begin{algorithmic}[1]
		\STATE \textbf{Initialize:} nominal $\{\mathbf{w}_k^{(0)}\}_{k\in\mathcal{K}}$, $\vect{v}^{(0)}$, $\eta_0$, iteration index $q=0$, time $t_0$ = 0, $t_\mathsf{out} = 0$
		\STATE Calculate auxiliary variables $\vect{\gamma}_\mathsf{PS}^{(0)}$, $\vect{u}_\mathsf{PS}^{(0)}$, $\vect{r}_\mathsf{PS}^{(0)}$ using \eqref{P2SINR}, \eqref{gradVP1}, and \eqref{eq:r_expl_eta}, respectively
\STATE \textbf{Rate adaption:} Update rates according to \eqref{FBL_rate}
		\REPEAT
		\STATE \textbf{Beamformer update:}
		\STATE Solve \eqref{P1.2} for $\{\mathbf{w}_k^{(q+1)},\vect{\gamma}_\mathsf{BF}^{(q+1)}, \vect{u}_\mathsf{BF}^{(q+1)}, \vect{r}_\mathsf{BF}^{(q+1)}\}$ with fixed $\vect{v}^{(q)}$, $\eta_q$, using $\vect{\gamma}_\mathsf{PS}^{(q)}$, $\vect{u}_\mathsf{PS}^{(q)}$, $\vect{r}_\mathsf{PS}^{(q)}$
		\STATE \textbf{Phase shifter update:}
		\STATE Solve \eqref{P2} for $\vect{v}^{(q+1)},\vect{\gamma}_\mathsf{PS}^{(q+1)}, \vect{u}_\mathsf{PS}^{(q+1)}, \vect{r}_\mathsf{PS}^{(q+1)}$ with fixed $\{\mathbf{w}_k^{(q+1)}\}$, $\eta_q$, using $\vect{\gamma}_\mathsf{BF}^{(q+1)}$, $\vect{u}_\mathsf{BF}^{(q+1)}$, $\vect{r}_\mathsf{BF}^{(q+1)}$
		\STATE \textbf{Blocklength update:}
		\STATE Compute $\eta_\text{obj}^\star$ with \eqref{eq:nobj} and bounds $\eta_k^\mathsf{Thr}$ with \eqref{eq:nthr}
		\STATE $\eta_{q+1} \leftarrow \left[\min \Big( \eta_\text{obj}^\star, \; \min_{k \in \mathcal{K}} \eta_k^\mathsf{Thr} \Big)\right]^{\eta_\mathsf{max}}_{\eta_\mathsf{min}},$
		\STATE $t_{q+1} \leftarrow t_q + \eta_{q+1}/B$ using \eqref{eq:TransTime}
		\STATE Update queues $\mat{Q}(t_{q+1})$  using \eqref{eq:virtQueue}
		\STATE $q \leftarrow q+1$
		\UNTIL{convergence: $\zeta \leq0$} or outage detected
		\STATE $t_\mathsf{abs} \leftarrow  t_q$
		\STATE $\{\vect{w}_k^\mathsf{abs}\} \leftarrow \vect{w}_k^{(q)} $, $\vect{v}^\mathsf{abs} \leftarrow \vect{v}^{(q)}$, $\eta^\mathsf{abs} \leftarrow \eta_q$
		\REPEAT
		\STATE \textbf{Beamformer update:}
		\STATE Solve \eqref{P5} for $\{\mathbf{w}_k^{(q+1)},\vect{\gamma}_\mathsf{BF}^{(q+1)}, \vect{u}_\mathsf{BF}^{(q+1)}, \vect{r}_\mathsf{BF}^{(q+1)}\}$ with fixed $\vect{v}^{(q)}$, $\eta_q$, using $\vect{\gamma}_\mathsf{PS}^{(q)}$, $\vect{u}_\mathsf{PS}^{(q)}$, $\vect{r}_\mathsf{PS}^{(q)}$
		\STATE \textbf{Phase shifter update:}
		\STATE Solve \eqref{P6} for $\vect{v}^{(q+1)},\vect{\gamma}_\mathsf{PS}^{(q+1)}, \vect{u}_\mathsf{PS}^{(q+1)}, \vect{r}_\mathsf{PS}^{(q+1)}$ with fixed $\{\mathbf{w}_k^{(i+1)}\}$, $\eta_q$, using $\vect{\gamma}_\mathsf{BF}^{(q+1)}$, $\vect{u}_\mathsf{BF}^{(q+1)}$, $\vect{r}_\mathsf{BF}^{(q+1)}$
		\STATE \textbf{Blocklength update:}
		\STATE Compute $\eta_\text{obj}^\star$ with \eqref{eq:nobj} and bounds $\eta_k^\mathsf{Thr}$ with \eqref{eq:nthr}
		\STATE $\eta_{q+1} \leftarrow \left[\min \Big( \eta_\text{obj}^\star, \; \min_{k \in \mathcal{K}} \eta_k^\mathsf{Thr} \Big)\right]^{\eta_\mathsf{max}}_{\eta_\mathsf{min}},$
		\STATE $t_{q+1} \leftarrow t_q + \eta_{q+1}/B$ using \eqref{eq:TransTime}
		\STATE Update queues $\mat{Q}(t_{q+1})$  using \eqref{eq:virtQueue}
		\STATE $q \leftarrow q+1$
		\UNTIL{convergence: $Q_k(t_q) \le Q_k^{\mathrm{des}},\, \forall k \in \mathcal{K}$} or outage detected
		\STATE $t_\mathsf{ada} \leftarrow t_q$
		\STATE Set $\{\vect{w}_k^\mathsf{abs}\} ,\vect{v}^\mathsf{abs}, \eta^\mathsf{abs}$ as new nominal values
	\end{algorithmic}
\end{algorithm}

\section{Numerical results}
In this section, we evaluate the performance of the proposed framework through numerical simulations under a realistic wireless network setting. The simulation setup considers a square area of $500 \times 500\,\text{m}^2$, where a single RIS, equipped with $M=900$ reflecting elements, is deployed at the center of the region. The \acp{AP} are randomly distributed within an annular region around the RIS to avoid unrealistically short propagation distances, with inner and outer radii of $50\,\text{m}$ and $330\,\text{m}$, respectively. A total of $K=30$ users are served and each \ac{AP} employs $10$ antennas. The system operates with a coherence time of $T_\mathsf{coh} = 3\,\text{ms}$ and a bandwidth of $B = 10\,\text{MHz}$, while the maximum transmit power per \ac{AP} is set to $P_n^\mathsf{max} = 32\,\text{dBm}$.

Unless stated otherwise the traffic model is based on stochastic packet arrivals, where an initial packet arrival realization is generated according to a Poisson process. Specifically, for each coherence interval \(T_{\mathrm{coh}}=3\,\text{ms}\), a single arrival realization is drawn using an arrival intensity of \(\xi_k=15000\) packets/s, resulting in an expected number of arrivals of \(\bar{A}_k=\xi_kT_{\mathrm{coh}}=45\) packets per interval. The obtained realization is then used to characterize the corresponding mean traffic rates for the queue model. Packets are assumed to have a fixed size of \(M_k^p=1536\) bits, allowing the queue dynamics to be equivalently represented in the bit domain. Consequently, the system can be interpreted as a discrete-time queue with stochastic traffic demand and rate-based service.

Based on the generated traffic realization, a corresponding minimum service rate is defined to ensure that the incoming data can be served within one coherence interval. Using the above definitions, the Poisson-generated packet arrivals result in user-specific minimum service rates \vspace{-0.1cm}
\[
\alpha_k^\mathsf{des} = \frac{A_k M_k^p}{T_{\mathrm{coh}}},
\]
which vary between users due to the stochastic traffic realization. For the considered setup, the resulting service requirements range from approximately \(18.9\) Mbps to \(30.7\) Mbps, with \(30.7\) Mbps representing the highest observed demand. The desired and maximum virtual queue thresholds are defined as \(Q_k^\mathsf{des}=0.25M_k^p\) and \(Q_k^\mathsf{max}=2M_k^p\), respectively. We recall that \(Q_k^\mathsf{max}\) represents the maximum tolerated virtual backlog deficit, meaning that the system cannot sustain a service deficit exceeding two packet equivalents due to insufficient allocated QoS. Finally, the target block error rate is set to \(\epsilon=10^{-3}\), the phase shift penalty constant is set to $\alpha_{v}=1000$ and the blocklength is constrained within \(\eta_{\min}=128\) and \(\eta_{\max}=512\), with initialization \(\eta_0=256\), reflecting the short-packet regime considered in this work.

In addition to the queue dynamics, we incorporate an outage model to capture severe channel degradation events. Within each coherence interval, an outage event is modeled by removing a direct \ac{AP}-user link from the system until the end of the coherence interval. The affected links are selected based on their channel strength, measured by \(\|\vect{h}_{n,k}\|_2^2\), where consecutive outage events progressively target stronger direct links, starting from the third strongest link and ending with the strongest channel realization. Only direct AP-user channels are subject to outages, while RIS-assisted channel components are assumed to remain available, thereby isolating the impact of direct-link failures on system performance. This modeling choice is motivated by the need to assess the resilience of the proposed framework under abrupt and highly non-uniform channel disruptions. By removing the strongest links, we intentionally stress the system to evaluate its ability to maintain service continuity and recover under adverse conditions.

\subsection{Temporal System Behavior Under Repeated Outage Events}

\subsubsection{Comparison with Benchmark Approaches}

\begin{figure}[!ht]
  \centering

  \begin{subfigure}{1\linewidth}
    \includegraphics[width=\linewidth]{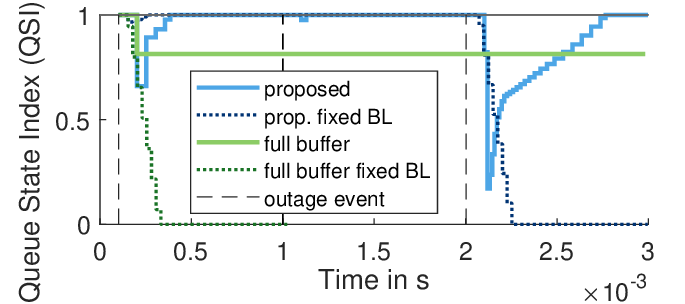}
    \caption{Queue State Index (QSI)}
  \end{subfigure}

  \begin{subfigure}{1\linewidth}
    \includegraphics[width=\linewidth]{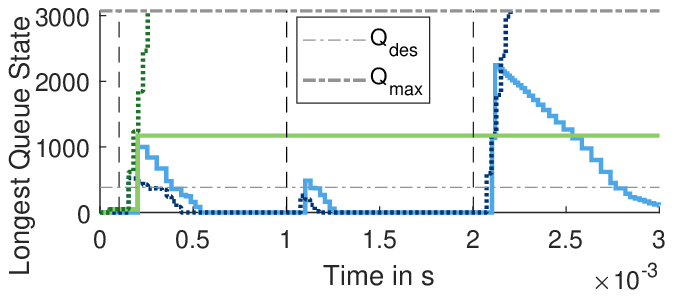}
    \caption{Longest queue state}
  \end{subfigure}

  \begin{subfigure}{1\linewidth}
    \includegraphics[width=\linewidth]{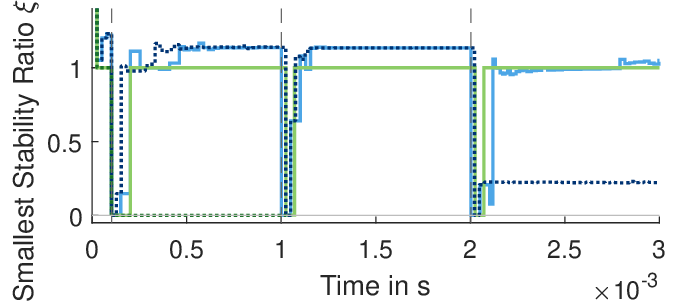}
    \caption{Smallest stability ratio}
  \end{subfigure}

    \begin{subfigure}{1\linewidth}
    \includegraphics[width=\linewidth]{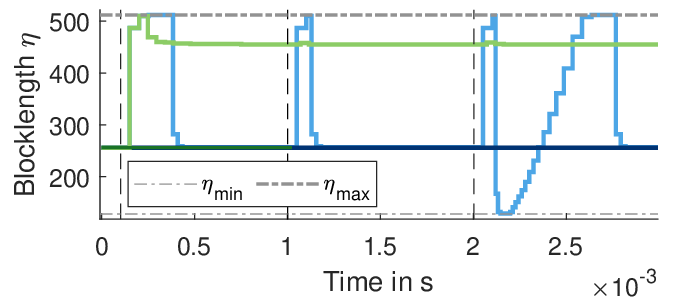}
    \caption{Blocklength}
  \end{subfigure}

\caption{Temporal evolution of the resilience behavior during consecutive outage events, illustrating how the absorption-adaptation mechanism jointly regulates the QSI (a), longest queue state (b), smallest stability ratio (c), and blocklength (d). The response of the proposed framework is compared against fixed-blocklength and full-buffer benchmark approaches.}
\label{fig:tempEv}

\end{figure}

Fig.~\ref{fig:tempEv} illustrates the temporal evolution of the proposed resilience framework during three consecutive outage events occurring at $t=\{0.2,\,1,\,2\}\,\mathrm{ms}$. At each outage, the strongest direct user--AP channel is removed, where the final outage therefore corresponds to the strongest channel in the Frobenius-norm sense. The figure depicts the evolution of the Queue State Index (QSI), the maximum queue state in the network, the minimum stability ratio, and the optimized blocklength in Fig.~\ref{fig:tempEv}(a)--(d), respectively. Four approaches are compared: the proposed Two-Phase Temporal Resilience Optimization with dynamic blocklength adaptation (solid bright blue), the proposed framework with fixed blocklengths (dotted dark blue), the resilience-oriented full-buffer baseline of~\cite{weinberger2025symbolcountsresilientwireless} (dotted dark green), and the baseline extended with dynamic blocklength optimization (solid bright green).

\textit{Proposed Two-Phase Temporal Resilience Optimization (solid bright blue):}
The proposed framework successfully absorbs and recovers from all three outage events. Following each disruption, the minimum stability ratio in Fig.~\ref{fig:tempEv}(c) immediately drops below one, indicating that the service rate of at least one user temporarily falls below its arrival rate. Consequently, the corresponding queue begins to accumulate packets, causing the maximum queue state in Fig.~\ref{fig:tempEv}(b) to increase beyond its desired value $Q_k^\mathrm{des}$ and the QSI in Fig.~\ref{fig:tempEv}(a) to decrease.

During the absorption phase, the optimizer increases the blocklength, as shown in Fig.~\ref{fig:tempEv}(d), thereby reducing the finite blocklength (FBL) rate penalty and progressively restoring the stability ratio towards $\xi_k \approx 1$. Once stable operation has been re-established, the framework enters the adaptation phase. In this regime, the transmission rates are intentionally increased beyond the desired operating point, yielding stability ratios larger than one and actively draining the accumulated queue backlog. As a result, all queues return below $Q_k^\mathrm{des}$, restoring the QSI to one.

The third outage represents the most challenging operating point since multiple strong channels have already been removed. Consequently, the minimum stability ratio remains much closer to one during recovery, indicating that only limited excess resources are available for backlog reduction. In this resource-constrained regime, blocklength adaptation becomes particularly important. The optimizer initially selects larger blocklengths to reduce the FBL penalty during absorptionand then decreases the blocklength during adaptation to optimize queue depletion. Interestingly, the optimal blocklength after each adaption step settles at progressively larger values, reflecting the challenging channel conditions while still enabling full recovery.

\textit{Proposed framework with fixed blocklengths (dotted dark blue):}
Removing the dynamic blocklength adaptation does not fundamentally alter the resilience mechanism. The framework successfully absorbs and recovers from the first two outage events and even exhibits slightly faster convergence because the fixed blocklength allow the SCA iterations to converges a little quicker. However, after the third outage the blocklength is no longer available to reduce the FBL penalty and free up additional resources. Consequently, the minimum stability ratio remains below one, preventing complete recovery of the affected queue. The queue eventually exceeds its maximum allowable length $Q_k^\mathrm{max}$, causing the QSI to collapse to zero.

\textit{Resilience-oriented full-buffer baseline (dotted dark green):}
The full-buffer baseline of~\cite{weinberger2025symbolcountsresilientwireless} fails already after the first outage event. Since resources cannot be reallocated from users without queue accumulation, the affected user never regains a stability ratio above one. As a result, its queue continuously grows until reaching $Q_k^\mathrm{max}$, permanently driving the QSI to zero.

\textit{Baseline with dynamic blocklength adaptation (solid bright green):}
Equipping the baseline with dynamic blocklength optimization substantially improves its robustness. Following each outage, the optimizer increases the blocklength until the minimum stability ratio approaches one, thereby preventing further queue growth and enabling the framework to withstand all three outage events. Effectively, the dynamic blocklength optimization enables the algorithm to satisfy the blocklength threshold identified in~\cite{weinberger2025symbolcountsresilientwireless} as necessary for obtaining a resilient system response. Nevertheless, unlike the proposed two-phase framework, the baseline only restores queue stability but does not actively eliminate the accumulated backlog. Consequently, the queues remain permanently above their desired operating point after the first outage, resulting in a persistently degraded QSI despite maintaining stable operation.

\subsubsection{Resilience Performance under Imperfect CSI}
To assess the performance of the proposed method under practical channel uncertainty, we also assume imperfect CSI at the CP. As described in Section~\ref{ssec:chan}, the MMSE-based channel estimation leads to Gaussian estimation errors modeled independently for the direct and RIS-assisted channels as
$
\Delta\vect h_k \sim \mathcal{CN}\!\left(\mathbf{0}, \sigma_{h,k}^2 \mathbf{I}\right), \quad
\Delta\mat G_k \sim \mathcal{CN}\!\left(\mathbf{0}, \sigma_{G,k}^2 \mathbf{I}\right),
$
where the error variances are assumed proportional to the respective channel powers as
$
\sigma_{h,k}^2 = \rho_e \,\mathbb{E}\!\left[\|\vect h_k\|_2^2\right], \quad
\sigma_{G,k}^2 = \rho_e \,\mathbb{E}\!\left[\|\mat G_k\|_F^2\right].
$
Here, $\rho_e =0.05$ denotes a relative channel estimation accuracy factor of 5\%. During the evaluation, we assume that the CP only has the estimates $\hat{\vect h}_k$ and $\hat{\mat G}_k$ available during the Two-Phase Temporal Resilience Optimization in Alg.~\ref{alg:alt_opt}. To isolate the impact of the proposed optimization framework, the pilot overhead required for channel estimation is neglected.

\begin{figure}[!t]
  \centering
  \includegraphics[width=1\linewidth]{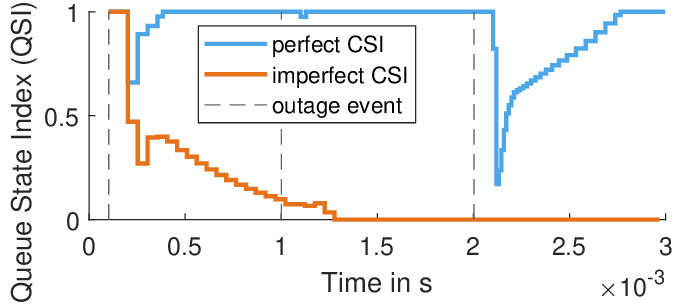}
\caption{Evolution of the QSI under ideal and imperfect CSI. Channel estimation errors reduce the achievable service rates, resulting in degraded recovery performance and eventual queue overflow under repeated outage events.}  \label{fig:estCSI}
\end{figure}

Figure~\ref{fig:estCSI} illustrates the evolution of the QSI when the central processor only has access to imperfect CSI. The orange curve corresponds to the proposed algorithm under the channel estimation error model described above, while the reference curve represents the ideal CSI case. Under imperfect CSI, the proposed framework is still able to initiate the absorption--adaptation mechanism after the first outage. However, compared with the ideal CSI case, a more pronounced degradation of the QSI can be observed. After the first outage event, the QSI decreases significantly and only partially recovers. This behavior is caused by the mismatch between the estimated and the actual channel conditions. Since the optimization is performed based on the estimated CSI, the selected transmission parameters lead to an overestimation of the achievable rates. Consequently, the actual service rates are lower than expected, resulting in a gradual accumulation of the queues despite the recovery action of the algorithm.

After the second outage event, this mismatch becomes critical. The additional channel degradation combined with the accumulated queue backlog prevents the system from fully restoring the desired queue states. As a result, the maximum queue constraint $Q_k^\mathrm{max}$ is violated and the QSI drops to zero.

The observed degradation highlights the importance of incorporating CSI uncertainty directly into the resilience optimization. A potential extension of the proposed framework could introduce robust or stochastic rate constraints, where the optimization accounts for channel estimation errors by including safety margins or probabilistic guarantees on the achievable rates. Such an approach would reduce the mismatch between predicted and actual service rates and improve resilience under practical channel uncertainty.

\subsection{Resilience Performance Under Bursty AI-Driven Traffic}
To evaluate the proposed resilience framework under AI-driven traffic surges, the packet size is modeled as an event-dependent quantity
\[
M_k^p(t_q)=M_0\beta_k^p(t_q),
\]
where \(M_0=256\) bits denotes the baseline packet size and \(\beta_k^p(t_q)\) represents the packet scaling factor of user \(k\) during AI traffic event \(t_q\). Initially, all users are assigned an identical nominal workload with
\[
\bar{\beta}_k^p=\beta_k^p(t_0)=3.
\]
Subsequently, at each AI-driven traffic spike event, the instantaneous packet scaling factors are drawn from a Poisson distribution with mean \(\bar{\beta}_k^p\), i.e.,
\[
\beta_k^p(t_q)\sim\mathrm{Poisson}(\bar{\beta}_k^p),
\]
thereby capturing the stochastic variations in the baseline AI workload across users.

On top of this heterogeneous baseline traffic, event-driven workload bursts are introduced by randomly activating additional high-demand users at each AI traffic event. At every successive traffic spike, a new user is selected and experiences an additional workload increase, resulting in a gradual accumulation of burst-generating users over time. This burst component is modeled by increasing the packet scaling factor of the activated users by a fixed increment of $14$, corresponding to an approximately \(5.7\)-fold increase in packet size relative to the nominal workload. Hence, the traffic evolves from a balanced workload distribution toward progressively concentrated AI-driven traffic surges. Accordingly, the queue thresholds are adjusted to reflect the increased packet sizes and prevent immediate failure states. Specifically, the desired and maximum virtual queue levels are defined as \(Q_k^\mathsf{des}=0.25(M_0\bar{\beta}_k^p)\) and \(Q_k^\mathsf{max}=6(M_0\bar{\beta}_k^p)\), respectively. This preserves the relative queue tolerance while accounting for the larger AI-generated data packets.

Furthermore, the required service rate becomes event-dependent due to the varying AI workload intensity. After each AI traffic event, the corresponding QoS requirement is updated according to
\begin{align}
\alpha_k^\mathsf{AI}(t_q)
&=
\xi_k\mathbb{E}\!\left[M_k^p(t_q)\right] =
\xi_k M_0\mathbb{E}\!\left[\beta_k^p(t_q)\right],
\end{align}
which represents the average bit arrival rate required to maintain queue stability under the current AI traffic conditions.
\begin{figure}[!ht]
  \centering

  \begin{subfigure}{1\linewidth}
    \includegraphics[width=\linewidth]{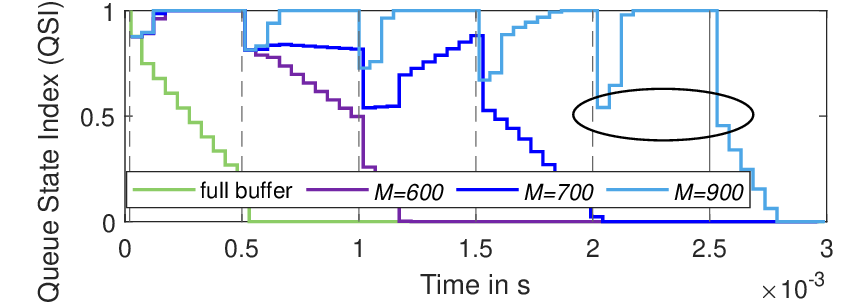}
    \caption{Queue State Index (QSI)}
  \end{subfigure}

  \begin{subfigure}{1\linewidth}
    \includegraphics[width=\linewidth]{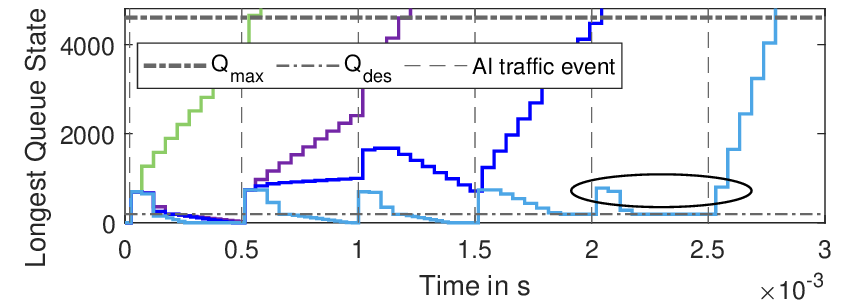}
    \caption{Longest queue state}
  \end{subfigure}
  \caption{Temporal evolution of the resilience behavior under AI-driven bursty traffic, illustrating the queue stability indicator (QSI) (a) and the longest queue state (b). The proposed framework is evaluated for different RIS sizes $M$ and compared with a full-buffer baseline with $M=900$ employing dynamic blocklength adaptation.}
  \label{fig:burstyTraff}
\end{figure}

Figure~\ref{fig:burstyTraff} illustrates the temporal evolution of the QSI in Fig.~\ref{fig:burstyTraff}(a) and the longest queue state in Fig.~\ref{fig:burstyTraff}(b) over successive AI traffic events. The proposed framework is evaluated for different \ac{RIS} sizes and compared with the full-buffer baseline of~\cite{weinberger2025symbolcountsresilientwireless}, equipped with dynamic blocklength adaptation and $M=900$ reflecting elements.

The full-buffer baseline is unable to absorb even the first traffic spike, causing a rapid accumulation of packets in the affected queue and resulting in system failure immediately after the second AI traffic event. In contrast, the proposed framework with only $M=600$ reflecting elements successfully absorbs the first traffic spike by dynamically reallocating communication resources from users with lightly loaded queues to users experiencing increased \ac{QoS} demands. Following the second traffic event, however, the available resources become insufficient, causing the queue states to gradually deteriorate until instability is reached.

Increasing the \ac{RIS} size to $M=700$ further improves the resilience of the proposed framework, as evidenced by the delayed onset of instability. Although the second traffic event again drives the system close to its resource limit, the proposed scheduling mechanism maintains balanced queue occupancies by redistributing resources among users, resulting in only a gradual degradation of both the QSI and the longest queue state. Interestingly, after the third traffic event, the framework successfully adapts to the modified traffic conditions and partially restores the queue states toward their nominal operating point. Only after the fourth traffic event do the increased \ac{QoS} requirements consistently exceed the available communication resources, ultimately causing system failure after the fifth event.

With $M=900$ reflecting elements, the proposed framework withstands and fully recovers from five consecutive AI traffic events before eventually reaching its resource limit after the sixth event. An interesting observation is that, although the longest queue state returns to nearly the same level after each successful recovery, the QSI exhibits a progressively larger degradation immediately following each traffic event, as highlighted by the ellipses in Fig.~\ref{fig:burstyTraff}(a) and (b). Since the QSI depends on the product of all queue states, this behavior indicates that the increased traffic demand of a single high-load user is redistributed across multiple users rather than being confined to the affected queue alone. This demonstrates that the proposed cross-layer framework balances the system-wide queue occupancy to preserve overall network stability. Compared with the full-buffer baseline, which lacks this cross-user resource adaptation capability, the proposed cross-layer resilience mechanism significantly improves the network's ability to absorb, adapt to, and recover from successive AI-driven traffic bursts.

\subsection{Resilience Metric}
\begin{figure}
  \centering
  \includegraphics[width=1\linewidth]{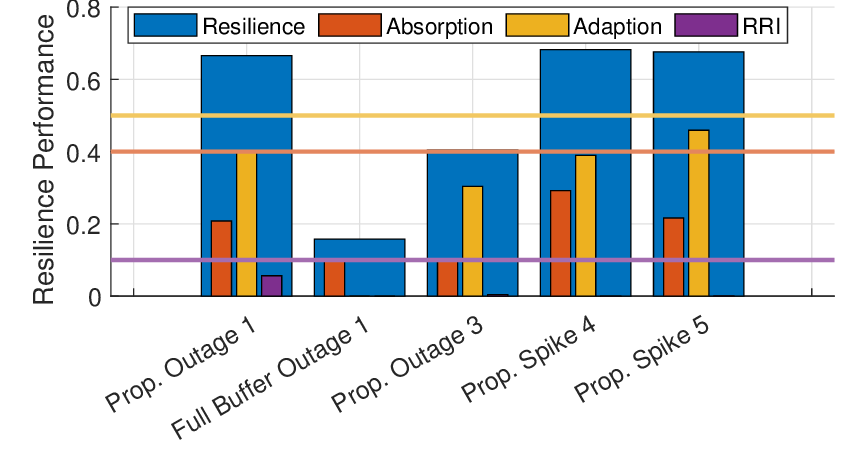}
  \caption{Resilience metric evaluation for different channel outage and AI traffic spike scenarios. The individual scaled contributions of the absorption, adaptation, and RRI components as well as their maximum values are shown together with the unified weighted resilience metric.}
  \label{fig:resMetric}
\end{figure}

To quantify the system behavior under different resilience events in an interpretable manner, this section evaluates the proposed resilience metric introduced in Section~\ref{sec:ResMetric}. The metric provides an intuitive grading of each event by jointly capturing the absorption capability, adaptation behavior, and recovery performance of the system. Figure~\ref{fig:resMetric} presents the resilience performance for five representative scenarios, including the first and third outage events of the proposed Two-Phase Temporal Resilience Optimization framework (corresponding to Fig.~\ref{fig:tempEv}), the first outage event of the full-buffer baseline Fig.~\ref{fig:tempEv}, and the fourth and fifth AI traffic spike events under bursty traffic conditions (corresponding to Fig.~\ref{fig:burstyTraff}).

The colored bars represent the individual $\lambda_i$-scaled resilience components, namely the absorption metric \(r_\mathsf{abs}\) \eqref{eq:r_abs}, adaptation metric \(r_\mathsf{ada}\) \eqref{eq:r_ada} , and Relative Recovery Index (RRI) \(r_\mathsf{rec}\) \eqref{eq:r_rec}. The overall resilience score in \eqref{eq:r} is obtained using the weighted combination
\(\lambda_1=0.4\), \(\lambda_2=0.5\), and \(\lambda_3=0.1\), for absorption, adaption and RRI, respectively, and is illustrated by the blue envelope surrounding each bar. The selected weighting prioritizes the absorption and adaptation capabilities, since maintaining queue stability under multiple unforeseen disruptions is the primary objective in the proposed URLLC system. Therefore, greater emphasis is placed on the ability of the system to initially absorb disruptions and subsequently adapt its resources to sustain stable operation under consecutive resilience events. The recovery speed, represented by the RRI, is assigned a lower weight, as a rapid recovery is beneficial but only meaningful if queue stability can first be preserved.

The first comparison considers the first channel outage event of the proposed framework and the corresponding full-buffer baseline (see Fig.~\ref{fig:tempEv}). As shown in Fig.~\ref{fig:resMetric}, the proposed framework successfully absorbs and adapts to the initial outage event. However, the severe rate degradation during the absorption phase results in a noticeable penalty in \(r_\mathsf{abs}\). In contrast, the adaptation component achieves a high score, as the proposed framework is able to fully restore the required QoS conditions. Nevertheless, the overall resilience score does not reach its maximum value, since both phases require a finite amount of time. In particular, the adaptation phase requires approximately twice the duration of the absorption phase, resulting in an RRI penalty of approximately \(0.5\).

In comparison, the full-buffer baseline is only able to absorb the initial outage but fails to adapt to the increased QoS requirements. Moreover, the absorption phase requires an additional time slot to restore all users above their required service rates, which further reduces the absorption score. Since the adaptation phase and recovery process are not achieved, both \(r_\mathsf{ada}\) and \(r_\mathsf{rec}\) are set to zero, resulting in a significantly lower overall resilience score.

Comparing the first and third channel outage events of the proposed framework in Fig.~\ref{fig:tempEv}, the latter represents a more challenging resilience event, as indicated by the increased degradation during absorption and the longer recovery duration. This is reflected by the reduced absorption and recovery-related scores. Nevertheless, since the framework successfully completes both absorption and adaptation phases, the resulting resilience score remains approximately three times higher than that of the full-buffer baseline.

The proposed framework is further evaluated under AI-driven traffic spikes in Fig.~\ref{fig:burstyTraff}, considering 4\textsuperscript{th} and 5\textsuperscript{th} spike events at \(1.5\)~ms and \(2\)~ms, respectively. Compared with channel outages, the absorption phase is significantly faster, as AI traffic spikes only require resource redistribution among users. In contrast, outage events require link recovery through RIS-assisted channel reconfiguration or alternative AP associations, resulting in more extensive changes to the resource allocation process. Consequently, \(r_\mathsf{abs}\) achieves higher values than in the outage scenarios. However, the more severe traffic increase at the \(2\)~ms spike results in a lower absorption score compared with the \(1.5\)~ms case. Conversely, the recovery from the more severe spike is faster, as fewer adaptation steps are required to restore the queue states, resulting in a comparable overall resilience score for both traffic events. Due to the rapid absorption and relatively longer adaptation phase, the corresponding RRI values remain lower.

Since the RRI provides insight into the balance between the absorption and adaptation phases, high values indicate a rapid transition from immediate reaction to long-term stabilization, whereas low values indicate that adaptation is the limiting factor. The consistently low RRI values observed across most resilience events show that the proposed framework effectively absorbs disturbances but operates with limited resources during adaptation. In contrast, the first outage achieves a higher RRI, as sufficient residual resources remain available after absorption, enabling faster stabilization. Hence, the RRI serves as an indicator of the remaining resource flexibility after absorption, revealing how effectively the available resources can be redirected toward adaptation and queue stabilization.

Overall, the proposed resilience metric enables not only a quantitative comparison between different resilience approaches but also a consistent evaluation across fundamentally different resilience events. By decomposing the performance into absorption, adaptation, and recovery behavior, the metric provides insight into the underlying mechanisms that determine resilience rather than only indicating whether the system eventually succeeds or fails. 

\looseness-1


\section{Conclusion}\label{ch:conc}
We proposed a unified resilience-oriented framework that jointly captures rate adaptation, queue dynamics, and blocklength optimization, enabling the system to react to and recover from severe channel disruptions and AI-driven traffic surges. In addition, we developed an interpretable resilience metric that provides a transparent evaluation of resilience events in terms of absorption, adaptation, and relative recovery behavior, allowing direct comparisons across different disruption types.

In doing so, we emphasized the importance of temporal effects in resilience, particularly in \ac{URLLC} scenarios where stability must be maintained under strict latency constraints. To capture these effects, we incorporated the finite blocklength regime, where transmission duration becomes an explicit design variable and provides additional degrees of freedom for time-domain optimization. This results in a two-stage operational structure consisting of an absorption phase, which prevents queue instability under sudden workload or channel degradations, and an adaptation phase, which reduces accumulated backlog and restores stable operation. Building on this structure, we developed a three-stage alternating optimization framework for each phase, jointly optimizing beamforming, phase shifts, and blocklength. This enables the derivation of tractable performance bounds and provides insights into the resilience-performance trade-offs.

Beyond the evaluation of the considered framework, the proposed resilience metric provides a step toward a more systematic assessment of resilient wireless systems. By enabling a consistent comparison across different resilience approaches and disruption types, it addresses the current lack of unified evaluation criteria and offers insights into which system components limit resilience performance. Consequently, such metrics can support future resilience-oriented designs by identifying suitable optimization targets and guiding the development of adaptive strategies for increasingly dynamic network conditions.
\section{Appendix}\label{chap:appendix}
\subsection{Analytical Solution for $\eta_q$}\label{app:A}
\noindent
To derive the optimal $\eta_q^*$, we first simplify $\bar{\delta}_k(\eta_q)$ by defining
\begin{align}
A_k &\triangleq C_k^q - \alpha_k^\mathsf{des}, \\
B_k &\triangleq B \Omega \sqrt{V_\mathsf{disp}(\Gamma_k^q)}, \\
D_k &\triangleq B(Q_k^\mathsf{max}-Q_k(t_q)).
\end{align}
Then
\begin{align}
\bar{\delta}_k(\eta_q) = \frac{-A_k \eta_q + B_k \sqrt{\eta_q}}{D_k}.
\end{align}

Introducing the substitution $\sqrt{\eta_q} = x$, we obtain a quadratic function:
\begin{align}
\bar{\delta}_k(x) = \frac{-A_k x^2 + B_k x}{D_k}.
\end{align}

Setting the derivative with respect to $x$ to zero gives the stationary point:
\begin{align}
\frac{d \bar{\delta}_k(x)}{dx} = \frac{-2 A_k x + B_k}{D_k} = 0 \quad \Rightarrow \quad x^* = \frac{B_k}{2 A_k}.
\end{align}

Reverting back to $\eta_q$, we obtain the candidate solution:
\begin{align}
\eta_{q,k}^\text{cand} = (x^*)^2 = \frac{B_k^2}{4 A_k^2} = \frac{(B \Omega)^2 V_\mathsf{disp}(\Gamma_k^q)}{4 (C_k^q - \alpha_k^\mathsf{des})^2}.
\end{align}

Finally, to satisfy the min--max objective and the feasible range $\eta_\mathsf{min} \le \eta_q \le \eta_\mathsf{max}$, the optimal blocklength becomes
\begin{align}
\eta_q^* = \left[ \max_{k\in\mathcal{K}} \frac{(B \Omega)^2 V_\mathsf{disp}(\Gamma_k^q)}{4 (C_k^q - \alpha_k^\mathsf{des})^2} \right]_{\eta_\mathsf{min}}^{\eta_\mathsf{max}}.
\end{align}
\subsection{Analytical Solution for $\eta^\star$ under FBL and Stability Constraint}\label{app:B}
\noindent
We derive the optimal blocklength $\eta^\star$ for the weighted backlog reduction objective under per-user stability constraint.

The objective is given by
\begin{align}
f(\eta) =  \sum_{k \in \mathcal{K}} \tilde Q_k \left[ (\alpha_k^\mathsf{des} - C_k) \eta + \Omega \sqrt{V_\mathsf{disp}} \sqrt{\eta} \right].
\end{align}

Introducing the substitution $x = \sqrt{\eta}$ yields a quadratic function in $x$:
\begin{align}
f(x) = \sum_{k \in \mathcal{K}} \tilde Q_k \big[ (\alpha_k^\mathsf{des} - C_k) x^2 + \Omega \sqrt{V_\mathsf{disp}} x \big].
\end{align}

Setting the derivative with respect to $x$ to zero gives the stationary point:
\begin{align}
\frac{d f(x)}{dx} = \sum_{k \in \mathcal{K}} \tilde Q_k \big[ 2 (\alpha_k^\mathsf{des} - C_k) x + \Omega u_k \big] = 0
\quad \nonumber \\\Rightarrow \quad
x^\star = \frac{\Omega \sum_{k \in \mathcal{K}} \tilde Q_k u_k}{-2 \sum_{k \in \mathcal{K}} \tilde Q_k (\alpha_k^\mathsf{des} - C_k)}.
\end{align}

Reverting back to $\eta$ gives the unconstrained FBL solution:
\begin{align}
\eta_\text{FBL}^\star = (x^\star)^2 =
\left(
\frac{\Omega \sum_{k \in \mathcal{K}} \tilde Q_k u_k}{-2 \sum_{k \in \mathcal{K}} \tilde Q_k (\alpha_k^\mathsf{des} - C_k)}
\right)^2.
\end{align}

The per-user stability constraint
\begin{align}
\frac{-\eta \, (r_k(\eta) - \alpha_k^\mathsf{des})}{B(Q_k^\mathsf{max}-Q_k)} \le \zeta^\mathsf{Thr}
\end{align}
can be rewritten as a quadratic in $x = \sqrt{\eta}$:
\begin{align}
(C_k - \alpha_k^\mathsf{des}) x^2 - B \, \Omega \, u_k \, x + \zeta^\mathsf{Thr} B (Q_k^\mathsf{max}-Q_k) \ge 0.
\end{align}

Solving this quadratic inequality for $x$ gives the feasible upper bound on $\sqrt{\eta}$:
\begin{align}
x \le \frac{B  \Omega  u_k - \sqrt{ (B  \Omega  u_k)^2 - 4 (C_k - \alpha_k^\mathsf{des})  \zeta^\mathsf{Thr} B (Q_k^\mathsf{max}-Q_k) }}{ 2 (C_k - \alpha_k^\mathsf{des}) }.
\end{align}

Hence, the per-user maximum blocklength allowed by the stability constraint is $\eta_k^\mathsf{Thr}  = x^2$, i.e.,
\begin{align}
\eta_k^\mathsf{Thr} &=
\biggl(
\frac{\vartheta_k - \sqrt{\vartheta_k^2 - 4 (C_k - \alpha_k^\mathsf{des}) \, \zeta^\mathsf{Thr} B (Q_k^\mathsf{max}-Q_k)}}
{2 (C_k - \alpha_k^\mathsf{des})}
\biggr)^2,
\end{align}
where $\vartheta_k = B \Omega u_k$.

\footnotesize
\bibliographystyle{IEEEtran}
\bibliography{references}

\begin{thebibliography}{10}
\providecommand{\url}[1]{#1}
\csname url@samestyle\endcsname
\providecommand{\newblock}{\relax}
\providecommand{\bibinfo}[2]{#2}
\providecommand{\BIBentrySTDinterwordspacing}{\spaceskip=0pt\relax}
\providecommand{\BIBentryALTinterwordstretchfactor}{4}
\providecommand{\BIBentryALTinterwordspacing}{\spaceskip=\fontdimen2\font plus
\BIBentryALTinterwordstretchfactor\fontdimen3\font minus
  \fontdimen4\font\relax}
\providecommand{\BIBforeignlanguage}[2]{{%
\expandafter\ifx\csname l@#1\endcsname\relax
\typeout{** WARNING: IEEEtran.bst: No hyphenation pattern has been}%
\typeout{** loaded for the language `#1'. Using the pattern for}%
\typeout{** the default language instead.}%
\else
\language=\csname l@#1\endcsname
\fi
#2}}
\providecommand{\BIBdecl}{\relax}
\BIBdecl

\bibitem{you2021towards}
X.~You, C.-X. Wang, J.~Huang, X.~Gao, Z.~Zhang, M.~Wang, Y.~Huang, C.~Zhang,
  Y.~Jiang, J.~Wang \emph{et~al.}, ``Towards {6G} wireless communication
  networks: Vision, enabling technologies, and new paradigm shifts,''
  \emph{Science China information sciences}, vol.~64, pp. 1--74, 2021.

\bibitem{ChaccourAOI}
C.~Chaccour and W.~Saad, ``On the ruin of age of information in augmented
  reality over wireless terahertz {(THz)} networks,'' in \emph{GLOBECOM}, 2020.

\bibitem{SemCom}
T.~M. Getu, G.~Kaddoum, and M.~Bennis, ``A survey on goal-oriented semantic
  communication: Techniques, challenges, and future directions,'' \emph{IEEE
  Access}, vol.~12, pp. 51\,223--51\,274, 2024.

\bibitem{brosInArms}
R.-J. Reifert, Y.~Karacora, C.~Chaccour, A.~Sezgin, and W.~Saad, ``Resilience
  and criticality: Brothers in arms for {6G},'' \emph{arXiv preprint
  arXiv:2412.03661}, 2024.

\bibitem{RobertRes}
R.-J. Reifert, S.~Roth, A.~A. Ahmad, and A.~Sezgin, ``Comeback kid: Resilience
  for mixed-critical wireless network resource management,'' \emph{IEEE Trans.
  on Vehicul. Techn}, vol.~72, no.~12, pp. 16\,177--16\,194, 2023.

\bibitem{ResByDesign}
N.~H. Mahmood, S.~Samarakoon, P.~Porambage, M.~Bennis, and M.~Latva-Aho,
  ``Resilient-by-design: A resilience framework for future wireless networks,''
  \emph{IEEE Commun. Mag.}, vol.~63, no.~11, pp. 158--164, 2025.

\bibitem{URLLC_Res}
W.~Saad, M.~Bennis, and M.~Chen, ``A vision of {6G} wireless systems:
  Applications, trends, technologies, and open research problems,'' \emph{IEEE
  Network}, vol.~34, no.~3, pp. 134--142, 2020.

\bibitem{weinberger2025symbolcountsresilientwireless}
K.~Weinberger and A.~Sezgin, ``When every symbol counts: Resilient wireless
  systems under finite blocklength constraints,'' in \emph{Europ. Wirel.},
  2025, pp. 46--51.

\bibitem{GenAI}
H.~Zou, Q.~Zhao, S.~Lasaulce, L.~Bariah, M.~Bennis, and M.~Debbah,
  ``{GenAINet}: Enabling wireless collective intelligence via knowledge
  transfer and reasoning,'' \emph{IEEE Access}, vol.~13, pp. 77\,764--77\,777,
  2025.

\bibitem{STERBENZ20101245}
J.~P. Sterbenz, D.~Hutchison, E.~K. {\c{C}}etinkaya, A.~Jabbar, J.~P. Rohrer,
  M.~Sch{\"o}ller, and P.~Smith, ``Resilience and survivability in
  communication networks: Strategies, principles, and survey of disciplines,''
  \emph{Computer networks}, vol.~54, no.~8, pp. 1245--1265, 2010.

\bibitem{resiliencemetric}
M.~Najarian and G.~J. Lim, ``Design and assessment methodology for system
  resilience metrics,'' \emph{Risk Anal.}, vol.~39, no.~9, pp. 1885--1898,
  2019.

\bibitem{Joint_BL_BF}
A.~A. Nasir, H.~D. Tuan, H.~H. Nguyen, M.~Debbah, and H.~V. Poor, ``Resource
  allocation and beamforming design in the short blocklength regime for
  {URLLC},'' \emph{IEEE Trans. on Wirel. Commun.}, vol.~20, no.~2, pp.
  1321--1335, 2021.

\bibitem{FBL_chanuse}
G.~Durisi, T.~Koch, and P.~Popovski, ``Toward massive, ultrareliable, and
  low-latency wireless communication with short packets,'' \emph{Proceedings of
  the IEEE}, vol. 104, no.~9, pp. 1711--1726, 2016.

\bibitem{basar2019wireless}
E.~Basar, M.~Di~Renzo, J.~De~Rosny, M.~Debbah, M.-S. Alouini, and R.~Zhang,
  ``Wireless communications through reconfigurable intelligent surfaces,''
  \emph{IEEE access}, vol.~7, pp. 116\,753--116\,773, 2019.

\bibitem{SynBenefits}
K.~Weinberger, A.~A. Ahmad, A.~Sezgin, and A.~Zappone, ``Synergistic benefits
  in {IRS}- and {RS}-enabled {C-RAN} with energy-efficient clustering,''
  \emph{IEEE Trans. on Wirel. Commun.}, 2022.

\bibitem{Dressler_Res}
A.~Zubow, I.~v. Stebut, S.~Rösler, and F.~Dressler, ``{ResCTC}: Resilience in
  wireless networks through cross-technology communication,'' in \emph{2024
  IEEE 35th PIMRC}, 2024, pp. 1--6.

\bibitem{bennis2025resilient}
M.~Bennis, S.~Samarakoon, T.~Alshammari, C.~Weeraddana, Z.~Tian, and C.~B.
  Issaid, ``Resilient-native and intelligent next-generation wireless systems:
  Key enablers, foundations, and applications,'' \emph{arXiv preprint
  arXiv:2506.22991}, 2025.

\bibitem{URLLC-FBL}
C.~She, C.~Yang, and T.~Q.~S. Quek, ``Cross-layer optimization for
  ultra-reliable and low-latency radio access networks,'' \emph{IEEE Trans. on
  Wirel. Commun.}, vol.~17, no.~1, pp. 127--141, 2018.

\bibitem{URLLC_RISOPT}
M.~Soleymani, I.~Santamaria, E.~A. Jorswieck, R.~Schober, and L.~Hanzo,
  ``Optimization of the downlink spectral- and energy- efficiency of
  {RIS}-aided multi-user {URLLC} {MIMO} systems,'' \emph{IEEE Trans. on
  Commun.}, vol.~73, no.~5, pp. 3497--3513, 2025.

\bibitem{RIS_RES}
K.~Weinberger, R.-J. Reifert, A.~Sezgin, and E.~Basar, ``{RIS}-enhanced
  resilience in cell-free {MIMO},'' in \emph{WSA and SCC}, 2023.

\bibitem{Joint_RIS_FBL_BF}
S.~Pala, K.~Singh, M.~Katwe, and C.-P. Li, ``Joint optimization of {URLLC}
  parameters and beamforming design for multi-{RIS}-aided {MU-MISO} {URLLC}
  system,'' \emph{IEEE Wirel. Commun. Lett.}, vol.~12, no.~1, pp. 148--152,
  2023.

\bibitem{channelEst}
Q.-U.-A. Nadeem, H.~Alwazani, A.~Kammoun, A.~Chaaban, M.~Debbah, and M.-S.
  Alouini, ``Intelligent reflecting surface-assisted multi-user {MISO}
  communication: Channel estimation and beamforming design,'' \emph{IEEE
  OJCOM}, vol.~1, pp. 661--680, 2020.

\bibitem{FBL_Polyanskiy}
Y.~Polyanskiy, H.~V. Poor, and S.~Verdu, ``Channel coding rate in the finite
  blocklength regime,'' \emph{IEEE Trans. on Inform. Theory}, vol.~56, no.~5,
  pp. 2307--2359, 2010.

\bibitem{neely2010introduction}
M.~J. Neely, ``Introduction to queues,'' in \emph{Stoch. Network Opt. with
  Appl. to Commun. and Queueing Systems}.\hskip 1em plus 0.5em minus
  0.4em\relax Springer, 2010, pp. 15--28.

\bibitem{accelRec}
K.~Weinberger, R.-J. Reifert, A.~Sezgin, and M.~Bennis, ``Accelerated recovery
  with {RIS}: Designing wireless resilience in mission-critical environments,''
  \emph{arXiv preprint arXiv:2504.11589}, 2025.

\bibitem{Cleckx_approx}
J.~Xu and B.~Clerckx, ``Max-min fairness and {PHY}-layer design of uplink
  {MIMO} rate-splitting multiple access with finite blocklength,'' \emph{IEEE
  Trans. on Commun.}, vol.~73, no.~5, pp. 3671--3682, 2025.

\end{thebibliography}

\begin{IEEEbiography}[{\includegraphics[width=1in,height=1.25in,clip,keepaspectratio]{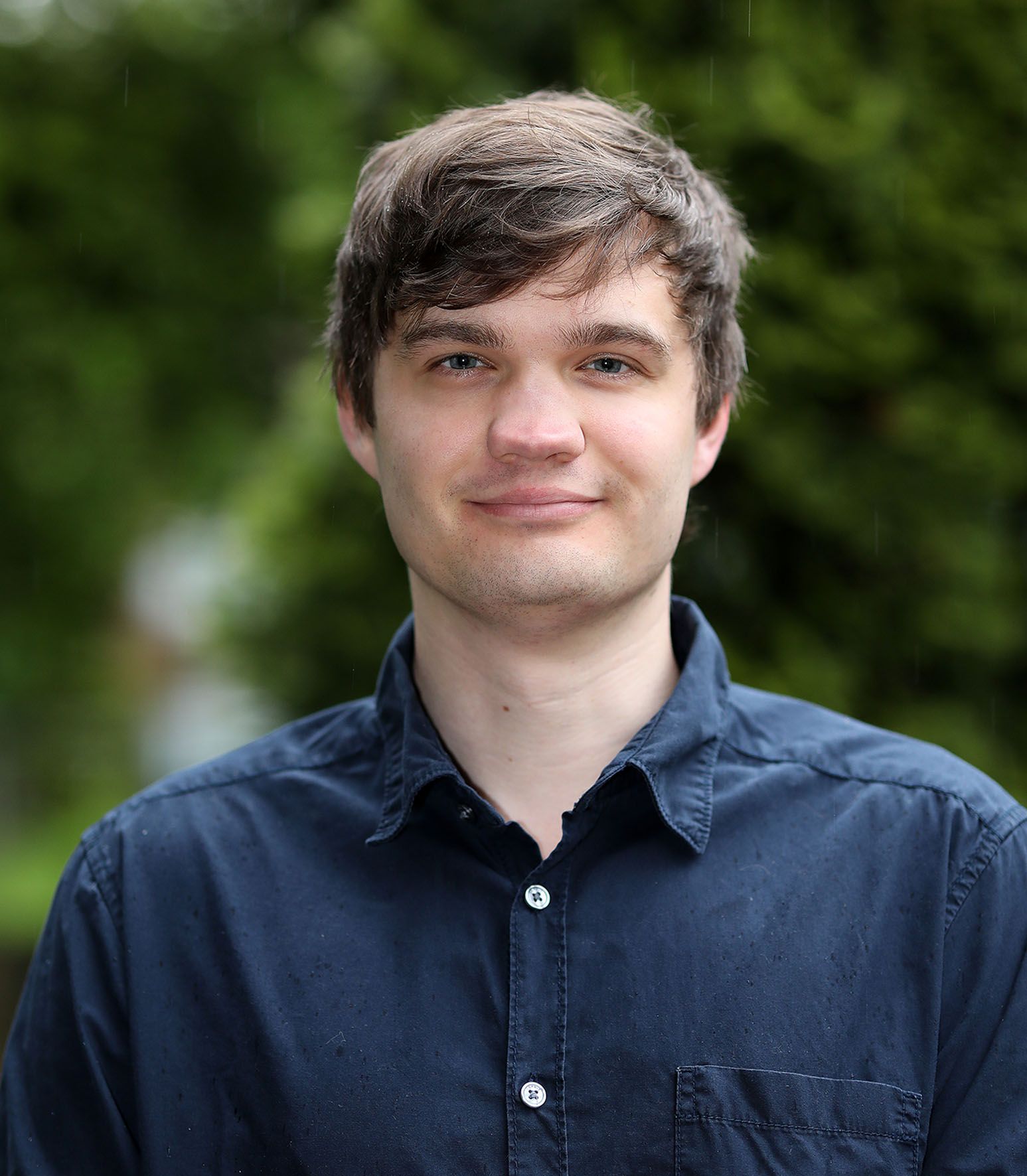}}]{Kevin Weinberger}
received the B.Sc. and M.Sc. degrees in electrical engineering and information technology from Ruhr University Bochum, Germany, in 2017 and 2020, respectively, where he is currently pursuing the Ph.D. degree with the Institute of Digital Communication Systems. His research interests include wireless communications, information theory, signal processing, and optimization for next-generation communication systems. His current research focuses on resilient wireless networks, reconfigurable intelligent surfaces, finite blocklength communications, and cross-layer resource optimization for 6G communication systems.
\end{IEEEbiography}
\begin{IEEEbiography}[{\includegraphics[width=1in,height=1.25in,clip,keepaspectratio]{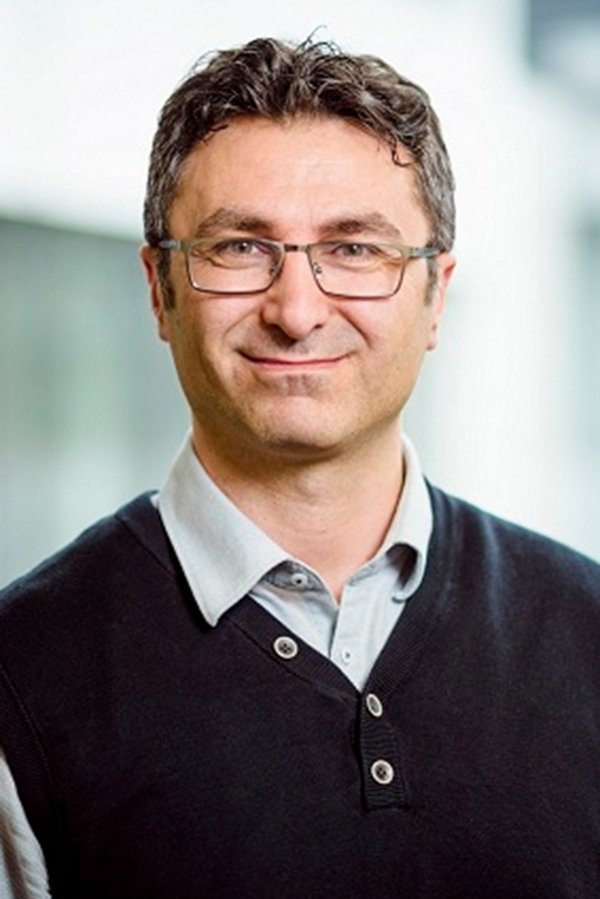}}]{Aydin Sezgin} (Senior Member, IEEE) received
the Dr. Ing. (Ph.D.) degree in electrical engineering from TU Berlin, in 2005. From 2001 to 2006,
he was with the Heinrich-Hertz-Institut, Berlin.
From 2006 to 2008, he held a postdoctoral position, and was also a Lecturer with the Information
Systems Laboratory, Department of Electrical
Engineering, Stanford University, Stanford, CA,
USA. From 2008 to 2009, he held a postdoctoral
position with the Department of Electrical Engineering and Computer Science, University of California, Irvine, CA, USA.
From 2009 to 2011, he was the Head of the Emmy-Noether-Research Group
on Wireless Networks, Ulm University. In 2011, he joined TU Darmstadt,
Germany, as a Professor. He is currently a Professor with Ruhr University
Bochum, Germany. He has published several book chapters, more than
70 journals and 200 conference papers in these topics. He is a winner of the
ITG-Sponsorship Award, in 2006. He was a First Recipient of the prestigious
Emmy-Noether Grant by the German Research Foundation in communication engineering, in 2009. He has co-authored articles that received the Best
Poster Award at the IEEE Communication Theory Workshop, in 2011, the
Best Paper Award at ICCSPA, in 2015, ICC, in 2019, and ISAP, in 2023.
\end{IEEEbiography}
\begin{IEEEbiography}[{\includegraphics[width=1in,height=1.25in,clip,keepaspectratio]{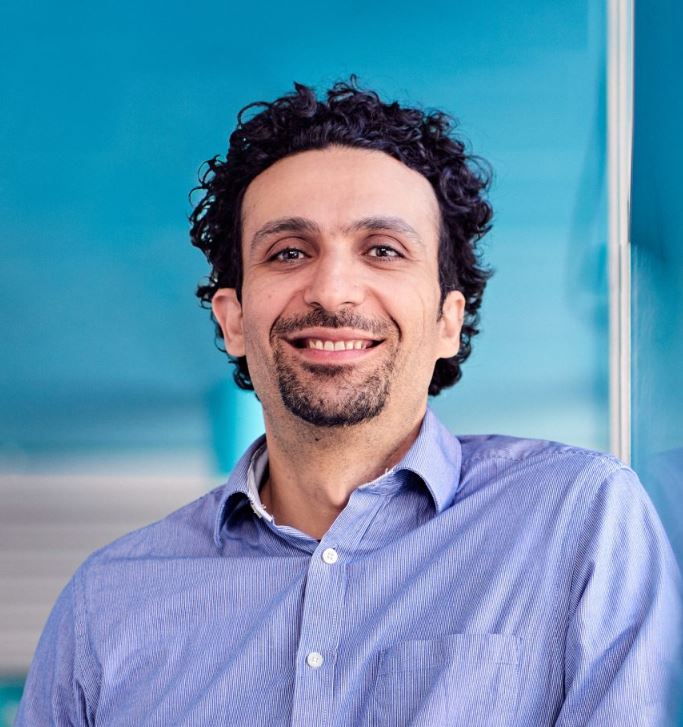}}]{Mehdi Bennis}
(Fellow, IEEE) is a Full (Tenured)
Professor with the Centre for Wireless Communications, University of Oulu, Oulu, Finland, and
the Head of the Intelligent Connectivity and Networks/Systems Group (ICON). His research interests include radio resource management, game
theory and distributed AI in 5G/6G networks. He
has published more than 300 research papers in
international conferences, journals and book chapters. He has been a recipient of several prestigious
awards including the 2015 Fred W. Ellersick Prize
from the IEEE Communications Society, the 2016 Best Tutorial Prize from
the IEEE Communications Society, the 2017 EURASIP Best paper Award
for the Journal of Wireless Communications and Networks, the all-University
of Oulu award for research, the 2019 IEEE ComSoc Radio Communications
Committee Early Achievement Award, and the 2020–2026 Clarviate Highly
Cited Researcher by the Web of Science.
\end{IEEEbiography}
\balance
\begin{acronym}
\setlength{\itemsep}{0.1em}
\acro{6G}{sixth-generation}
\acro{AF}{amplify-and-forward}
\acro{AI}{artificial intelligence}
\acro{AO}{alternating optimization}
\acro{AP}{access point}
\acro{AWGN}{additive white Gaussian noise}
\acro{B5G}{Beyond 5G}
\acro{BLER}{Block Error Rate}
\acro{bps}{Bits per second}
\acro{BS}{base station}
\acro{CB}{coherence block}
\acro{CE}{channel estimation}
\acro{C-RAN}{Cloud Radio Access Network}
\acro{CMD}{common message decoding}
\acro{CP}{central processor}
\acro{CSI}{channel state information}
\acro{CRLB}{Cramér-Rao lower bound}
\acro{D2D}{device-to-device}
\acro{DC}{difference-of-convex}
\acro{DFT}{discrete Fourier transformation}
\acro{DL}{downlink}
\acro{FBL}{finite blocklength}
\acro{GDoF}{generalized degrees of freedom}
\acro{IBL}{infinite blocklength}
\acro{IC}{interference channel}
\acro{i.i.d.}{independent and identically distributed}
\acro{IRS}{intelligent reflecting surface}
\acro{IoT}{Internet of Things}
\acro{LoS}{line-of-sight}
\acro{LSF}{large scale fading}
\acro{KPI}{key performance indicator}
\acro{M2M}{Machine to Machine}
\acro{MISO}{multiple-input and single-output}
\acro{MIMO}{multiple-input and multiple-output}
\acro{MRT}{maximum ratio transmission}
\acro{MRC}{maximum ratio combining}
\acro{MSE}{mean square error}
\acro{NOMA}{non-orthogonal multiple access}
\acro{NLoS}{non-line-of-sight}
\acro{PSD}{positive semidefinite}
\acro{QCQP}{quadratically constrained quadratic programming}
\acro{QoS}{quality-of-service}
\acro{RF}{radio frequency}
\acro{RC}{reflect coefficient}
\acro{RIS}{reconfigurable intelligent surface}
\acro{RS-CMD}{rate splitting and common message decoding}
\acro{RSMA}{rate-splitting multiple access}
\acro{RS}{rate splitting}
\acro{SCA}{successive convex approximation}
\acro{SDP}{semidefinite programming}
\acro{SDR}{semidefinite relaxation}
\acro{SIC}{successive interference cancellation}
\acro{SINR}{signal-to-interference-plus-noise ratio}
\acro{SOCP}{second-order cone program}
\acro{SVD}{singular value decomposition }
\acro{TTR}{time-to-recovery}
\acro{TIN}{treating interference as noise}
\acro{TDD}{time-division duplexing}
\acro{TSM}{topological signal management}
\acro{UHDV}{Ultra High Definition Video}
\acro{UL}{uplink}
\acro{URLLC}{ultra-reliable low-latency communication}
\acro{w.r.t.}{with respect to}
\acro{QoS}{Quality-of-Service}

\acro{AF}{amplify-and-forward}
\acro{AWGN}{additive white Gaussian noise}
\acro{B5G}{Beyond 5G}
\acro{BS}{base station}
\acro{C-RAN}{Cloud Radio Access Network}
\acro{CSI}{channel state information}
\acro{CMD}{common-message-decoding}
\acro{CM}{common-message}
\acro{CoMP}{coordinated multi-point}
\acro{CP}{central processor}
\acro{D2D}{device-to-device}
\acro{DC}{difference-of-convex}
\acro{EE}{energy efficiency}
\acro{IC}{interference channel}
\acro{i.i.d.}{independent and identically distributed}
\acro{IRS}{intelligent reflecting surface}
\acro{IoT}{Internet of Things}
\acro{LoS}{line-of-sight}
\acro{LoSC}{level of supportive connectivity}
\acro{M2M}{Machine to Machine}
\acro{NOMA}{non-orthogonal multiple access}
\acro{MISO}{multiple-input and single-output}
\acro{MIMO}{multiple-input and multiple-output}
\acro{MMSE}{minimum mean squared error}
\acro{MRT}{maximum ratio transmission}
\acro{MRC}{maximum ratio combining}
\acro{NLoS}{non-line-of-sight}
\acro{PA}{power amplifier}
\acro{PSD}{positive semidefinite}
\acro{QCQP}{quadratically constrained quadratic programming}
\acro{QoS}{quality-of-service}
\acro{RF}{radio frequency}
\acro{RRU}{remote radio unit}
\acro{RS-CMD}{rate splitting and common message decoding}
\acro{RS}{rate splitting}
\acro{SDP}{semidefinite programming}
\acro{SDR}{semidefinite relaxation}
\acro{SIC}{successive interference cancellation}
\acro{SCA}{successive convex approximation}
\acro{SINR}{signal-to-interference-plus-noise ratio}
\acro{SOCP}{second-order cone program}
\acro{SVD}{singular value decomposition }
\acro{TP}{transition point}
\acro{TIN}{treating interference as noise}
\acro{UHDV}{Ultra High Definition Video}
\acro{LoSC}{level of supportive connectivity}
\end{acronym}

\end{document}